\documentclass[11pt,a4paper]{article}

\usepackage{bm}
\usepackage[top=15truemm,bottom=20truemm,left=20truemm,right=20truemm]{geometry}
\usepackage{fancyhdr}
\fancypagestyle{firstpage}{
  \fancyhf{}
  
  \fancyhead[R]{TTI-MATHPHYS-46} 
}
\fancypagestyle{nonfirstpage}{
  \fancyhf{}  
   
 \fancyfoot[C]{\hyperlink{contents}{\thepage}} 
}
\usepackage[colorlinks=true,urlcolor=blue,anchorcolor=black,citecolor=blue,linkcolor=black,filecolor=black,menucolor=black,linktocpage=true,pdfproducer=medialab,pdfa=true]{hyperref}
\usepackage{graphicx}
\usepackage{cite}
\usepackage{amsmath,latexsym,amssymb,mathrsfs,ascmac,physics,mathtools,slashed,xcolor}
\usepackage{comment}
\numberwithin{equation}{section}
\allowdisplaybreaks[1]
\newtheorem{theorem}{Theorem}[section]
\newcommand{\proofend}{\unskip\nobreak\hfill\ensuremath{\square}}
\makeatletter
\newcommand{\equationalias}[1]{%
  \begingroup
  \protected@edef\@currentlabel{\theequation}%
  \label{#1}%
  \endgroup}
\makeatother
\usepackage[affil-it,noblocks]{authblk}
\newcommand{\email}[1]{\thanks{\href{mailto:#1}{\texttt{#1}}}}

\usepackage{booktabs,needspace}

\makeatletter
\newcommand{\equationrefalias}[2]{%
  \begingroup
  \protected@edef\@currentlabel{\getrefnumber{#1}}%
  \protected@edef\@currentHref{\getrefbykeydefault{#1}{anchor}{}}%
  \label{#2}%
  \endgroup}
\makeatother

\title{A separation criterion for chiral systems\\ and its application to Pleba\'nski--Demia\'nski perturbations}

\author[1]{Hideo Furugori
\email{hideo@toyota-ti.ac.jp}}
\affil[1]{Mathematical Physics Laboratory, Toyota Technological Institute, Nagoya 468-8511, Japan}

\author[2]{Daisuke Yoshida\email{yoshida.daisuke.k9@f.mail.nagoya-u.ac.jp}}
\affil[2]{Institute for Advanced Research, Nagoya University, Nagoya 464-8602, Japan}
\date{}

\begin{document}

\maketitle
\thispagestyle{firstpage}

\begin{abstract}
We give a constructive criterion for local separation of
first-order chiral systems coupling two complex doublets.
It combines a Gram decomposition of the coupling sources
with separate coordinate dependence and a differential
transport test.
For recoverable, nondegenerate source families, the criterion
is necessary and sufficient within the specified class of
mode-independent complex matrix transformations with unit
determinant.
For a fixed differential operator and admissible Gram
decomposition, we identify conditions under which the sources
uniquely determine the Hermitian coefficient matrix required
for separation.
As an application, we construct an explicit algebraic
transformation and derive separated radial and angular
equations for coupled radiative Einstein--Maxwell
perturbations of rotating, charged Pleba\'nski--Demia\'nski
backgrounds with acceleration, NUT charge, and a cosmological
constant, on specified local patches.
Its Gram decomposition is unique under additional generic
assumptions.
We also derive angular endpoint indices for Bonnor and
Misner identifications with conical deficits or excesses.
Under the stated regularity assumptions, each specified
self-adjoint angular realization at fixed real mode
parameters has a discrete spectrum and a complete
orthonormal eigenbasis.
\end{abstract}

\setcounter{tocdepth}{2}
\hypertarget{contents}{\tableofcontents}

\pagestyle{nonfirstpage}

\section{Introduction}

Separation of variables provides a way to study perturbations of
rotating black holes through radial and angular ordinary differential
equations. For Kerr spacetime, Teukolsky obtained separable equations
for gravitational and electromagnetic perturbations
\cite{Teukolsky:1972my,Teukolsky:1973ha}.
On a charged background, the metric and the electromagnetic field
that supports it must be perturbed together, giving a coupled system.
The separation problem asks whether this coupled dynamics can be
organized into radial and angular systems; it does not require the
gravitational and electromagnetic components within each system to
decouple. A complementary tensorial formulation for Kerr--Newman
was developed in Ref.~\cite{Giorgi:2020ujd}.

Recent work has established such a separation for the coupled radiative
equations on Kerr--Newman and Kerr--Newman--(anti-)de Sitter backgrounds.
Hintz constructed a matrix integrating factor for a chiral first-order
formulation of the Kerr--Newman system~\cite{Hintz:2026jiw}.
Subsequent developments include numerical quasinormal mode calculations
\cite{Destounis:2026ljd}, static tidal responses
\cite{Chakraborty:2026kyf}, and a formulation regular in the zero-charge
limit that identifies physical radiative amplitudes
\cite{Motohashi:2026wac}.
The separation was extended to Kerr--Newman--de Sitter in
Ref.~\cite{Yoshida:2026yly}, which also analyzed angular completeness
for real frequencies and the relation to physical horizon conditions.
Zhao obtained a compact algebraic separating matrix for
Kerr--Newman--(anti-)de Sitter~\cite{Zhao:2026gen}.
Using an alternative algebraic separating matrix with an explicit
$\Lambda\to0$ limit, Sarkar et al.\ computed gravitoelectromagnetic
quasinormal modes of Kerr--Newman--de Sitter black holes and investigated
their implications for strong cosmic censorship~\cite{Sarkar:2026xkq}.

These matrix constructions lead to a general question about
separation of chiral first-order systems.
For a pair of complex doublets with adjugate-related coupling
matrices, which properties of the coupling matrix and the
Hermitian coefficient matrix permit a common, mode-independent
transformation to one-variable equations?
We address this structural question independently of the
system's physical origin and give a procedure for constructing
the transformation when the conditions are satisfied.

Our first main result is a constructive separation criterion
for this class of chiral systems
(Theorem~\ref{thm:complete-criterion}).
The determinant equips the real vector space of Hermitian
$2\times2$ matrices with a Lorentzian metric.
An additive decomposition of the source Gram matrix with
the required ranks, signatures, and separate coordinate
dependence determines two orthogonal internal planes.
A source frame adapted to these planes gives an algebraic
transformation that separates the coupling.
A further transport test determines whether it also separates
the differential equations.
For pointwise independent Hermitian sources recoverable from
the mode family, these conditions are necessary and sufficient
for local separation within the specified class of
mode-independent $SL(2,\mathbb C)$ transformations.
For a fixed differential operator and admissible Gram
decomposition, we also identify conditions under which the
coupling sources uniquely determine the compatible Hermitian
coefficient matrix
(Theorem~\ref{thm:hermitian-transport}).
When this uniqueness holds, the remaining radial boost and
angular rotation cannot accommodate an arbitrary choice of
that coefficient matrix.
The transport test can be performed on the source frame
before reconstructing the separating matrix.

We also ask whether the structure underlying the
Kerr--Newman--(anti-)de Sitter separation persists in the presence
of acceleration and NUT charge.
The Pleba\'nski--Demia\'nski (PD) family provides a setting in which
to test this structure, incorporating rotation, acceleration, NUT,
electric and magnetic charges, and a cosmological constant
\cite{Plebanski:1976gy,Griffiths:2005qp}.
We use its rotating sector in the explicit physical parametrization
of Refs.~\cite{Podolsky:2022xxd,Podolsky:2023erratum}.
Related separability results include massless field equations on the
vacuum spinning C metric~\cite{Bini:2008mzd} and Maxwell test fields
on PD~\cite{Frolov:2018pys}.
The latter propagate on a fixed geometry, whereas perturbations of
the electromagnetic field supporting a charged background also
couple to metric perturbations. Test-field separation therefore does
not by itself establish separation of the coupled radiative system.

Our second main result is an explicit algebraic matrix
$\mathsf X$, given in Eq.~\eqref{eq:pd-Xa}, that separates
the coupled gravitational and electromagnetic radiative
equations on rotating, charged PD backgrounds.
It yields the two-component radial and angular systems
in Eqs.~\eqref{eq:pd-Xa-radial} and \eqref{eq:pd-Xa-angular},
linked by a common separation constant.
The matrix is independent of the Fourier frequency and
azimuthal number and applies to both signs of spin weight,
including for complex frequencies.
We obtain it by verifying the general criterion for the PD
chiral system and constructing a separated source frame.
The construction is local on the regular patches specified
below and retains all background parameters without an expansion.
The Gram decomposition is unique under the additional generic
assumptions of Appendix~\ref{app:pd-rank-comparison};
these assumptions are not required for the explicit construction.
Its Kerr--Newman and Kerr--Newman--de Sitter limits recover
the earlier separations through the normalizations and
frame changes described in Sec.~\ref{sec:pd-special-backgrounds}.

Acceleration and NUT charge also introduce global angular features
absent from the usual regular-axis Kerr--Newman--de Sitter problem.
The azimuthal period controls conical deficits or excesses, while
NUT charge requires a choice of time identification.
We distinguish the nonperiodic-time Bonnor interpretation
\cite{Bonnor:1969} from the patched, periodic-time Misner interpretation
\cite{Misner:1963}, and derive their mode labels and endpoint indices.
Under the stated angular regularity assumptions and at fixed
real mode parameters, every specified self-adjoint angular
realization has compact resolvent and a complete orthonormal
eigenbasis.
Limit-point endpoints require no additional boundary condition,
whereas limit-circle endpoints require a choice of self-adjoint
domain.
For existing solutions with the required domain and regularity
properties, the angular eigenfunction expansion gives the
separated radial equations for its coefficients.
Local separation itself neither selects the angular domain
nor proves reconstruction of arbitrary spacetime perturbations.

This approach is related in motivation to separation through matrix
symmetry operators for Dirac equations
\cite{Kalnins:1986Dirac,Shapovalov:2017Dirac}, but starts from coupling
sources and their invariant bilinear products.
Appendix~\ref{app:other-systems} gives explicit reductions of a
translation-reduced Dirac system and the Fu--Kane model at zero
chemical potential~\cite{Fu:2008Kane} to the adjugate chiral form.
The Dirac example also contains a simple separable subclass.
Their source spans have dimension at most three, so these examples
distinguish membership in the chiral class from applicability
of the nondegenerate necessity and uniqueness results.

Sec.~\ref{sec:separation-structure} develops the general criterion,
summarized in Theorem~\ref{thm:complete-criterion}.
Secs.~\ref{sec:background} and \ref{sec:separated-equations}
formulate the PD system, verify the criterion, and construct
$\mathsf X$ and the separated equations.
Sec.~\ref{sec:pd-global-angular} treats global identifications,
conicity, angular endpoint conditions, and spectral completeness.
Sec.~\ref{sec:summary} summarizes the results and discusses
further applications.
Appendix~\ref{app:pd-rank-comparison} establishes generic uniqueness
of the PD Gram decomposition, and Appendix~\ref{app:other-systems}
presents the two examples outside gravitational perturbation theory.

\textbf{Note added.} While this manuscript was being finalized,
Gray and Ovcharenko posted a related work on arXiv~\cite{Gray:2026rzu},
reporting separation of the coupled gravitational and electromagnetic
perturbation equations on Pleba\'nski--Demia\'nski backgrounds.
The present work develops a constructive criterion based on a decomposition of the source Gram matrix and a transport test, establishes necessity and uniqueness results under specified assumptions, and analyzes the angular consequences of global identifications and conicity.

\section{Separation mechanism for chiral systems}
\label{sec:separation-structure}
The main objective of this paper is to separate the coupled gravitational
and electromagnetic radiative equations on the PD background.
Before treating the background explicitly, we isolate the mathematical
structure underlying separation of the chiral first-order system
and formulate it in terms of general theorems.
This formulation identifies which properties of the coupling matrix and
the Hermitian coefficient matrix make separation possible and provides a
systematic way to construct a separating transformation.
The PD application in Sec.~\ref{sec:separated-equations} will then
reduce to verifying these properties for the given coefficients.

We begin by specifying a separated form and showing how it yields
ordinary differential equations in each variable.
We then determine how to construct a transformation that separates
the coupling and how to verify the transport identity and
the separation of the transport function.
The results culminate in Theorem~\ref{thm:complete-criterion}, which,
under the stated source assumptions, gives necessary and sufficient
conditions for local separation within the specified class of
mode-independent transformations.

\subsection{Chiral equations and the separation theorem}
\label{sec:general-separation-theorem}
On a smooth product patch with normalized coordinates $x,y$,
consider the chiral pair
\begin{equation}
 (\mathscr D-\mathsf A)y_+=\mathsf F(\mathfrak m)y_-,\qquad
 (\overline{\mathscr D}+\mathsf A)y_-=\mathsf F(\mathfrak m)^\#y_+,
 \qquad \mathsf A^\dagger=\mathsf A,\quad\operatorname{tr}\mathsf A=0,
 \label{eq:general-chiral-system}
\end{equation}
\equationrefalias{eq:general-chiral-system}{eq:general-A-conditions}
where
\begin{equation}
 \mathscr D\coloneqq \partial_x-i\partial_y,\qquad
 \overline{\mathscr D}\coloneqq \partial_x+i\partial_y.
 \label{eq:general-chiral-derivatives}
\end{equation}
Here $\mathfrak m$ denotes the mode labels, and $y_\pm$ are complex
two-component columns, so one chiral pair contains four scalar
first-order equations.
The matrix $\mathsf A$ is mode independent.
For any complex $2\times2$ matrix $\mathsf M$,
we denote its adjugate by $\mathsf M^\#$.
The Cayley--Hamilton theorem gives
\begin{equation}
 \mathsf M^\#=(\operatorname{tr}\mathsf M)\mathsf I-\mathsf M,\qquad
 \mathsf M\mathsf M^\#=(\det \mathsf M)\mathsf I,\qquad
 \tfrac12\operatorname{tr}(\mathsf M^\#\mathsf M)=\det \mathsf M.
 \label{eq:adjugate-definition}
\end{equation}
These identities also hold at singular $\mathsf M$.

The Pauli matrices in the standard representation are
\begin{equation}
 \sigma_1\coloneqq \begin{pmatrix}0&1\\1&0\end{pmatrix},\quad
 \sigma_2\coloneqq \begin{pmatrix}0&-i\\i&0\end{pmatrix},\quad
 \sigma_3\coloneqq \begin{pmatrix}1&0\\0&-1\end{pmatrix}.
 \label{eq:direct-pauli}
\end{equation}
To allow an arbitrary constant choice of basis, let
$\mathsf U_{\mathsf T}\coloneqq (e_+,e_-)$ be any constant unitary matrix and set
$\tau_i\coloneqq \mathsf U_{\mathsf T}\sigma_i\mathsf U_{\mathsf T}^\dagger$ and
$\mathsf T\coloneqq \tau_3$.
The matrices $\tau_i$ are Hermitian and satisfy the Pauli algebra
\begin{equation}
 \tau_i\tau_j=\delta_{ij}\mathsf I+i\epsilon_{ijk}\tau_k.
 \label{eq:adapted-pauli-algebra}
\end{equation}
In particular, $\mathsf T$ is Hermitian, and
$\mathsf T e_\pm=\pm e_\pm$.

We seek a change of fields that allows separation into functions
of $x$ and $y$.
For a smooth, mode-independent $\mathsf X\in SL(2,\mathbb C)$, define
\begin{equation}
 \psi_+\coloneqq \mathsf X^{-1}y_+,\qquad \psi_-\coloneqq \mathsf X^\dagger y_-,\qquad
 \mathsf C\coloneqq \mathsf X^{-1}\mathsf F (\mathsf X^\dagger)^{-1}.
 \label{eq:direct-field-transformation}
\end{equation}
\equationrefalias{eq:direct-field-transformation}{eq:general-field-transformation}
The following theorem gives conditions on $\mathsf X$ under which
a paired product ansatz reduces the chiral equations to ordinary
differential equations in $x$ and $y$, linked by a single separation
constant.
\begin{theorem}[Separation of a chiral pair]
\label{thm:chiral-separation}
Suppose $\mathsf X$ satisfies the following three conditions.
\begin{enumerate}
\item Transport identity.
\begin{equation}
 \mathscr D\mathsf X=\mathsf A\mathsf X+\chi\mathsf X\mathsf T.
 \label{eq:general-X-transport}
\end{equation}
\equationrefalias{eq:general-X-transport}{eq:direct-X-transport}
Here $\chi$ is a real, mode-independent scalar function,
which we call the transport function.

\item Separation of the transport function.
\begin{equation}
 \chi(x,y)=\xi(x)+\eta(y).
 \label{eq:general-transport-function-separation}
\end{equation}

\item Separation of the coupling.
\begin{equation}
 \mathsf C=c_0(x;\mathfrak m)\mathsf I+c_3(x;\mathfrak m)\mathsf T
       +c_1(y;\mathfrak m)\tau_1+c_2(y;\mathfrak m)\tau_2.
 \label{eq:general-coupling-target}
\end{equation}
\equationrefalias{eq:general-coupling-target}{eq:direct-coupling-target}
The coefficients $c_i$ may be complex.
\end{enumerate}
For fixed mode labels, consider fields of the form
\begin{align}
 \psi_+&=R_1(x)S_1(y)e_++R_2(x)S_2(y)e_-,\nonumber\\
 \psi_-&=R_2(x)S_1(y)e_++R_1(x)S_2(y)e_-
 \label{eq:sep-ansatz}
\end{align}
\equationrefalias{eq:sep-ansatz}{eq:general-paired-ansatz}
Here we write
\begin{equation}
 R(x)\coloneqq \begin{pmatrix}R_1(x)\\R_2(x)\end{pmatrix},
 \qquad
 S(y)\coloneqq \begin{pmatrix}S_1(y)\\S_2(y)\end{pmatrix}.
 \label{eq:sep-amplitude-columns}
\end{equation}
We assume $R\not\equiv0$ and $S\not\equiv0$.
Under these assumptions, these fields solve the transformed chiral equations if and only if, for a constant
$\zeta$,
\begin{align}
 \mathcal L S&\coloneqq\begin{pmatrix}
 -i\partial_y+\eta&-c_1+i c_2\\ c_1+i c_2&i\partial_y+\eta
 \end{pmatrix}S=\zeta S,
 \label{eq:sep-angular-equation}\\
 \partial_xR
 &=\begin{pmatrix}
 -\zeta-\xi&c_0+c_3\\c_0-c_3&\zeta+\xi
 \end{pmatrix}R.
 \label{eq:sep-radial-equation}
\end{align}
\equationrefalias{eq:sep-radial-equation}{eq:general-radial-equation}
\equationrefalias{eq:sep-angular-equation}{eq:general-angular-equation}
\equationrefalias{eq:sep-angular-equation}{eq:sep-angular-operator}
\equationrefalias{eq:sep-angular-equation}{eq:general-angular-operator}
\end{theorem}

\paragraph{Proof.}
Differentiating $\mathsf X\mathsf X^{-1}=\mathsf I$ and using
the transport identity~\eqref{eq:general-X-transport} gives
\begin{equation}
 \mathscr D(\mathsf X^{-1})=-\mathsf X^{-1}(\mathscr D\mathsf X)\mathsf X^{-1}
 =-\mathsf X^{-1}\mathsf A-\chi \mathsf T \mathsf X^{-1}.
 \label{eq:inverse-transport}
\end{equation}
Applying this identity to $\psi_+=\mathsf X^{-1}y_+$
and using the first chiral equation gives
\begin{equation}
 \begin{aligned}
 \mathscr D\psi_+
 &=(\mathscr D\mathsf X^{-1})y_+
   +\mathsf X^{-1}\mathscr D y_+
 =-\chi\mathsf T\psi_++\mathsf C\psi_-.
 \end{aligned}
 \label{eq:sep-plus-derivation}
\end{equation}
Since $\mathsf A$ and $\mathsf T$ are Hermitian and $\chi$ is real,
$\overline{\mathscr D}\mathsf X^\dagger=\mathsf X^\dagger \mathsf A+\chi \mathsf T\mathsf X^\dagger$.
The adjugate reverses products and $\det \mathsf X=1$, hence
$\mathsf C^\#=\mathsf X^\dagger \mathsf F^\#\mathsf X$.
Applying these identities to $\psi_-=\mathsf X^\dagger y_-$
and using the second chiral equation gives
\begin{equation}
 \overline{\mathscr D}\psi_-
 =(\overline{\mathscr D}\mathsf X^\dagger)y_-
  +\mathsf X^\dagger\overline{\mathscr D}y_-
 =\chi\mathsf T\psi_-+\mathsf C^\#\psi_+.
 \label{eq:sep-minus-derivation}
\end{equation}
Thus the transformed system is
\begin{equation}
 (\mathscr D+\chi \mathsf T)\psi_+=\mathsf C\psi_-,\qquad
 (\overline{\mathscr D}-\chi \mathsf T)\psi_-=\mathsf C^\#\psi_+.
 \label{eq:sep-transformed-system}
\end{equation}
\equationrefalias{eq:sep-transformed-system}{eq:general-transformed-system}
Substituting the paired ansatz~\eqref{eq:sep-ansatz} gives
\begin{equation}
 \begin{aligned}
 \bigl(\partial_xR_1+\xi R_1-(c_0+c_3)R_2\bigr)S
 &=-R_1\mathcal LS,\\
 \bigl(\partial_xR_2-\xi R_2-(c_0-c_3)R_1\bigr)S
 &=R_2\mathcal LS.
 \end{aligned}
 \label{eq:sep-product-identities}
\end{equation}
Fixing any $x$ at which $R(x)\ne0$ yields
$\mathcal LS=\zeta S$ for a constant $\zeta$.
Substitution then gives the radial equation
\eqref{eq:sep-radial-equation}, since $S$ is not identically zero.
Conversely, the separated equations imply the transformed
chiral system.
\proofend

\paragraph{Changes of frame.}
The separating matrix is not unique. If $\mathsf X$ satisfies
the three conditions of Theorem~\ref{thm:chiral-separation},
then so does
\begin{equation}
  \mathsf X'= \mathsf X e^{\left(f(x)+ig(y)\right)\mathsf T},
 \qquad f,g\ \text{real}.
 \label{eq:separation-residual-gauge}
\end{equation}
Here $f$ and $g$ are arbitrary smooth, real, mode-independent
functions of their respective variables. The coupling
coefficients transform as
\begin{equation}
 (c_0\pm c_3)'=e^{\mp2f}(c_0\pm c_3),\qquad
 (c_1\mp i c_2)'=e^{\mp2ig}(c_1\mp i c_2),
 \label{eq:separated-coupling-frame-action}
\end{equation}
while the transport function remains additively separated, with
\begin{equation}
 \xi'=\xi+\partial_x f,\qquad
 \eta'=\eta+\partial_y g.
 \label{eq:normalization-proof-transport-gauge}
\end{equation}
The corresponding separated amplitudes are
\begin{equation}
 R_{\rm new}\coloneqq e^{-f(x)\sigma_3}R,\qquad
 S'\coloneqq \mathsf U_gS,\qquad \mathsf U_g(y)\coloneqq e^{-ig(y)\sigma_3}.
 \label{eq:residual-amplitude-action}
\end{equation}
Substitution into the paired ansatz gives the transformed
fields associated with $\mathsf X'$ and leaves the original
fields $y_\pm$ unchanged. The angular expressions satisfy
\begin{equation}
 \mathcal L'=\mathsf U_g\mathcal L \mathsf U_g^{-1}.
 \label{eq:angular-frame-equivalence}
\end{equation}
Local choices $\partial_x f=-\xi$ and
$\partial_y g=-\eta$ therefore give a representative
with vanishing transport function.
A different normalization may nevertheless yield simpler
explicit matrix entries.

Separately, the decomposition of
$\chi=\xi+\eta$ has the freedom
$\xi\mapsto\xi+c$, $\eta\mapsto\eta-c$.
The separated equations are unchanged if
$\zeta\mapsto\zeta-c$ at the same time.

\subsection{Constructing a transformation that separates the coupling}
\label{sec:coupling-construction}

Theorem~\ref{thm:chiral-separation} gives separated equations
once a matrix $\mathsf X$ satisfying its three conditions is available.
We now address the construction of such a matrix, beginning with
the separation of the coupling $\mathsf F$.
In the same smooth local setting, we show how a decomposition
of the source Gram matrix yields a transformation satisfying
the coupling condition.
The transport identity and the separation of the transport
function will be examined in
Sec.~\ref{sec:differential-compatibility}.

Assume a source representation
\begin{equation}
  \mathsf F(x,y;\mathfrak m)=
 \sum_{a=1}^{4}d_a(\mathfrak m) \mathsf S_a(x,y),
 \qquad  \mathsf S_a^\dagger= \mathsf S_a,
 \label{eq:general-source-family}
\end{equation}
where the $ \mathsf S_a$ are independent of the modes and form a real
basis of $\operatorname{Herm}(2)$ at each point.
The coefficients $d_a$ may be complex but are independent of $x,y$.

To obtain the coupling separation in
Eq.~\eqref{eq:general-coupling-target}, it is sufficient
to find a common transformation of the sources such that
the coefficients of $\mathsf I,\mathsf T$ depend only on $x$,
while those of $\tau_1,\tau_2$ depend only on $y$.
Since the coefficients $d_a$ are independent of $x,y$,
the same separation then holds for their combination $\mathsf F$.

On $\operatorname{Herm}(2)$, we use the real bilinear form
\begin{equation}
 \langle \mathsf U, \mathsf V\rangle
 \coloneqq\frac12\operatorname{tr}
 \bigl(\mathsf U^\# \mathsf V\bigr),
 \qquad
 \langle \mathsf U, \mathsf U\rangle=\det \mathsf U.
 \label{eq:separation-inner-product}
\end{equation}
\equationalias{eq:separation-proof-inner-product}
For $\mathsf U=u_0\mathsf I+u_i\sigma_i$ with real $u_\mu$,
Eq.~\eqref{eq:adjugate-definition} gives
$\mathsf U^\#=u_0\mathsf I-u_i\sigma_i$. Since
$\operatorname{tr}(\sigma_i\sigma_j)=2\delta_{ij}$,
we obtain
\begin{equation}
 \tfrac12\operatorname{tr}(\mathsf U^\#\mathsf V)=u_0v_0-\sum_{i=1}^3u_iv_i.
 \label{eq:separation-inner-product-components}
\end{equation}
Thus the bilinear form is real and symmetric, with signature $(+---)$.
For $\det\mathsf X=1$, the adjugate reverses products and
the cyclicity of the trace gives
\begin{equation}
 \begin{aligned}
 \langle\mathsf X\mathsf U\mathsf X^\dagger,
        \mathsf X\mathsf V\mathsf X^\dagger\rangle
 &=\tfrac12\operatorname{tr}
 \bigl((\mathsf X^\dagger)^{-1}
       \mathsf U^\#\mathsf V\mathsf X^\dagger\bigr)=\langle\mathsf U,\mathsf V\rangle.
 \end{aligned}
 \label{eq:separation-inner-product-invariance}
\end{equation}
If the transformed sources have this separated form,
their bilinear products split into an $x$-dependent part
and a $y$-dependent part.
There are no mixed terms because
$\langle\mathsf I,\tau_j\rangle
=\langle\mathsf T,\tau_j\rangle=0$ for $j=1,2$.
Equation~\eqref{eq:separation-inner-product-invariance}
shows that these products can be computed from the original
sources, without knowing $\mathsf X$.

We therefore collect them in the source Gram matrix,
\begin{equation}
 \Gamma_{ab}\coloneqq\langle\mathsf S_a,\mathsf S_b\rangle,
 \label{eq:general-source-gram}
\end{equation}
and seek a decomposition
\begin{equation}
 \Gamma(x,y)=\Gamma_x(x)+\Gamma_y(y),
 \qquad
 \operatorname{rank}\Gamma_x
 =\operatorname{rank}\Gamma_y=2,
 \label{eq:general-gram-split}
\end{equation}
where $\Gamma_x,\Gamma_y$ are real symmetric matrices.

For a decomposition induced by separated source coefficients, each part has rank at most two. The source basis assumption makes \(\Gamma\) nonsingular, so both ranks must be exactly two.

Their nonzero signatures are $(+-)$ and $(--)$,
respectively, because the $x$-dependent coefficients
multiply $\mathsf I,\mathsf T$, while the $y$-dependent
coefficients multiply $\tau_1,\tau_2$.
Any two additive decompositions differ by opposite
constant symmetric shifts:
\begin{equation}
 \Gamma_x\mapsto\Gamma_x+\mathsf K_0,\qquad
 \Gamma_y\mapsto\Gamma_y-\mathsf K_0,\qquad \mathsf K_0^T=\mathsf K_0.
 \label{eq:general-constant-split-freedom}
\end{equation}
We choose $\mathsf K_0$ so that both matrices have rank two
and the required signatures throughout the patch.

We have seen that separation of all the sources implies a Gram
decomposition with the rank and signature properties stated above.
The following theorem establishes the converse: from such a
decomposition and the sources, we can construct locally a
mode-independent matrix $\mathsf X$ satisfying the coupling separation
condition of Theorem~\ref{thm:chiral-separation}.
\begin{theorem}[Coupling separation]
\label{thm:coupling-construction}
Suppose that the following conditions hold.
\begin{enumerate}
\item
The coupling has the source representation
\eqref{eq:general-source-family}, where the smooth Hermitian
matrices $\mathsf S_a$ are mode-independent and form a real basis
of $\operatorname{Herm}(2)$ at each point.
The coefficients $d_a(\mathfrak m)$ are independent of $x,y$.

\item
The source Gram matrix defined in
Eq.~\eqref{eq:general-source-gram} admits a decomposition
$\Gamma(x,y)=\Gamma_x(x)+\Gamma_y(y)$ into smooth real symmetric
matrices of rank two.
The nonzero signatures of $\Gamma_x$ and $\Gamma_y$ are
$(+-)$ and $(--)$, respectively.
\end{enumerate}
Then, locally, there exists
a smooth mode-independent matrix
$ \mathsf X\in SL(2,\mathbb C)$
satisfying the coupling separation condition
\eqref{eq:general-coupling-target} of Theorem~\ref{thm:chiral-separation}. This matrix can be constructed from the sources and
the specified Gram decomposition alone; no differential operator
or matrix $ \mathsf A$ is required.
\end{theorem}

\paragraph{Proof.}
Locally choose smooth real $2\times4$ factors
$ \mathsf R(x)$ and $ \mathsf Q(y)$ such that
\begin{equation}
 \Gamma_x= \mathsf R^T
 \begin{pmatrix}1&0\\0&-1\end{pmatrix} \mathsf R,
 \qquad
 \Gamma_y=- \mathsf Q^T\mathsf Q.
 \label{eq:general-gram-factors}
\end{equation}
Such factors exist locally by the constant rank and signature
assumptions. The factors $\mathsf R$ and $\mathsf Q$ depend only
on $x$ and $y$, respectively.

Write the rows as $ \mathsf R_0, \mathsf R_1, \mathsf Q_0, \mathsf Q_1$,
and define
\begin{equation}
  \mathsf V\coloneqq
 \begin{pmatrix} \mathsf R_0\\ \mathsf R_1\\ \mathsf Q_0\\ \mathsf Q_1\end{pmatrix},
 \qquad
  \mathsf B_i\coloneqq
 \sum_{a=1}^4 \mathsf S_a(\mathsf V^{-1})_{a,i+1},
 \quad i=0,1,2,3.
 \label{eq:general-source-frame}
\end{equation}
The identity
$\Gamma=\mathsf V^T\operatorname{diag}(1,-1,-1,-1)\mathsf V$
makes $\mathsf V$ invertible.
Each $\mathsf B_i$ is Hermitian because it is a real linear
combination of the Hermitian source matrices $\mathsf S_a$.
We refer to $\mathsf B_i$ as the frame matrices. They satisfy
\begin{equation}
 \bigl(\langle \mathsf B_i,\mathsf B_j\rangle\bigr)_{ij}
 =\mathsf V^{-\mathsf T}\Gamma\mathsf V^{-1}
 =\operatorname{diag}(1,-1,-1,-1).
 \label{eq:general-source-frame-normalization}
\end{equation}
Thus the spans of $\mathsf B_0,\mathsf B_1$ and
$\mathsf B_2,\mathsf B_3$ are orthogonal nondegenerate
two-dimensional subspaces, which we call the radial and angular
planes, respectively.
These names anticipate the PD application in
Sec.~\ref{sec:separated-equations}, where $x$ and $y$
correspond to radial and angular variables, respectively.
For fixed sources, specifying $\Gamma_x$ and $\Gamma_y$
uniquely determines these planes, independently of the choice
of factors $\mathsf R$ and $\mathsf Q$.

To obtain coupling separation, it is enough to construct a common
matrix $\mathsf X\in SL(2,\mathbb C)$ such that
$\mathsf X^{-1}\mathsf B_i(\mathsf X^\dagger)^{-1}$
equals $\mathsf I,\mathsf T,\tau_1,\tau_2$
for $i=0,1,2,3$, respectively.
The factors $\mathsf R(x)$ and $\mathsf Q(y)$ then give the required
one-variable coefficients of the transformed sources.
We introduce the following auxiliary matrices to construct
$\mathsf X$.
Since $\mathsf B_0$ is Hermitian with determinant one,
it is either positive or negative definite.
Using the freedom in the choice of $\mathsf R$,
we take $\mathsf B_0>0$ and set
\begin{equation}
  \mathsf W\coloneqq  \mathsf B_0,\qquad
  \mathsf J\coloneqq  \mathsf B_1 \mathsf B_0^{-1},\qquad
  \mathsf E\coloneqq \mathsf B_3,\qquad
  \mathsf L\coloneqq  \mathsf E \mathsf W^{-1}.
 \label{eq:general-frame-matrices}
\end{equation}
We now verify the following algebraic relations:
\begin{equation}
 \begin{gathered}
 \operatorname{tr} \mathsf J=0,\qquad  \mathsf J^2= \mathsf I,\qquad
  \mathsf W= \mathsf W^\dagger>0,\qquad\det \mathsf W=1,\\
  \mathsf E= \mathsf E^\dagger,\qquad\det \mathsf E=-1,\qquad
  \mathsf J \mathsf W \mathsf J^\dagger= \mathsf W,\qquad
  \mathsf J \mathsf E \mathsf J^\dagger=- \mathsf E.
 \end{gathered}
 \label{eq:matrix-algebraic}
\end{equation}
\equationalias{eq:matrix-congruence}
\equationalias{eq:normalization-proof-algebraic}
The Hermiticity of $\mathsf W$ and $\mathsf E$ follows from
their definitions in Eq.~\eqref{eq:general-frame-matrices}
and the Hermiticity of the frame matrices.
Positivity of $\mathsf W$ was fixed above.
The determinant relations follow from the frame normalization
and $\langle\mathsf U,\mathsf U\rangle=\det\mathsf U$
in Eq.~\eqref{eq:separation-inner-product}.
To derive the first two relations in Eq.~\eqref{eq:matrix-algebraic},
we use $\mathsf W^{-1}=\mathsf W^\#$, which follows from
$\det\mathsf W=1$.
Orthogonality and normalization of the frame give
\begin{equation}
 \operatorname{tr}\mathsf J
 =\operatorname{tr}(\mathsf B_1\mathsf W^\#)
 =2\langle\mathsf W,\mathsf B_1\rangle=0,
 \qquad
 \det\mathsf J=\frac{\det\mathsf B_1}{\det\mathsf W}=-1.
 \label{eq:general-J-trace-determinant}
\end{equation}
The Cayley--Hamilton identity
$\mathsf J^2-(\operatorname{tr}\mathsf J)\mathsf J
+(\det\mathsf J)\mathsf I=0$
then yields $\mathsf J^2=\mathsf I$.
The penultimate relation in Eq.~\eqref{eq:matrix-algebraic}
follows from
$\mathsf J\mathsf W=\mathsf B_1=\mathsf W\mathsf J^\dagger$:
\begin{equation}
 \mathsf J\mathsf W\mathsf J^\dagger
 =\mathsf J^2\mathsf W=\mathsf W.
 \label{eq:general-JW-congruence}
\end{equation}

To derive the last relation in Eq.~\eqref{eq:matrix-algebraic},
introduce
\begin{equation}
 \widehat{ \mathsf J}\coloneqq  \mathsf W^{-1/2} \mathsf J \mathsf W^{1/2},
 \qquad
 \widehat{ \mathsf E}\coloneqq  \mathsf W^{-1/2} \mathsf E \mathsf W^{-1/2}.
 \label{eq:normalization-proof-whitened}
\end{equation}
The transformation $\mathsf M\mapsto\mathsf W^{-1/2}\mathsf M\mathsf W^{-1/2}$ preserves the bilinear form
and sends $\mathsf B_0$ to $\mathsf I$.
Thus $\widehat{\mathsf J}$ and $\widehat{\mathsf E}$ are
orthogonal traceless Hermitian matrices.
For any traceless $2\times2$ matrices $\mathsf U,\mathsf V$,
\begin{equation}
 \mathsf U\mathsf V+\mathsf V\mathsf U
 =\operatorname{tr}(\mathsf U\mathsf V)\mathsf I
 =-2\langle\mathsf U,\mathsf V\rangle\mathsf I.
 \label{eq:traceless-anticommutator}
\end{equation}
Applying this identity to the orthogonal matrices
$\widehat{\mathsf J}$ and $\widehat{\mathsf E}$
shows that they anticommute.
Since $\widehat{\mathsf J}^{\,2}=\mathsf I$, we obtain
\begin{equation}
 \widehat{\mathsf J}\widehat{\mathsf E}\widehat{\mathsf J}
 =-\widehat{\mathsf E}\widehat{\mathsf J}^{\,2}
 =-\widehat{\mathsf E}.
 \label{eq:general-normalized-congruence}
\end{equation}
Returning to the original matrices gives
$\mathsf J\mathsf E\mathsf J^\dagger=-\mathsf E$.

For the construction of $\mathsf X$ below,
we now establish three further relations:
$\mathsf L^2=\mathsf I$, $\{\mathsf J,\mathsf L\}=0$, and
$\mathsf B_2=\pm i\mathsf J\mathsf E$.
Since $\widehat{\mathsf E}$ is traceless with determinant $-1$,
the Cayley--Hamilton identity gives
$\widehat{\mathsf E}^{\,2}=\mathsf I$.
Together with the anticommutation of
$\widehat{\mathsf J}$ and $\widehat{\mathsf E}$,
this yields the first two relations by similarity transformation.
For the third relation,
$\mathsf W^{-1/2}\mathsf B_2\mathsf W^{-1/2}$ and
$i\widehat{\mathsf J}\widehat{\mathsf E}$
both lie in the one-dimensional orthogonal complement of
$\widehat{\mathsf J}$ and $\widehat{\mathsf E}$
within the space of traceless Hermitian matrices.
Both have determinant $-1$, so they agree up to sign.
Transforming back gives $\mathsf B_2=\pm i\mathsf J\mathsf E$.
Using the sign freedom in the choice of the row $\mathsf Q_0$,
we choose it so that the corresponding frame matrix satisfies
$\mathsf B_2=i\mathsf J\mathsf E$.

Using the columns $e_\pm$ of the fixed unitary matrix
$\mathsf U_{\mathsf T}$ used to define $\tau_i$, we construct
\begin{equation}
  \mathsf X\coloneqq
 \frac{\mathsf B}{\sqrt{\det\mathsf B}},
 \label{eq:pd-X-reconstruction}
\end{equation}
where
\begin{equation}
  \mathsf B\coloneqq
 (\mathsf I+\mathsf J)\beta e_+^\dagger
 +i(\mathsf I-\mathsf J)\beta e_-^\dagger,
 \label{eq:normalization-proof-column-matrix}
\end{equation}
and $\beta$ is any locally nonzero column of $ \mathsf W- \mathsf E$
and a smooth local square root is chosen.
To verify that the matrix $\mathsf X$ in
Eq.~\eqref{eq:pd-X-reconstruction} is well defined and satisfies
the required relations, note that in the frame defined by Eq.~\eqref{eq:normalization-proof-whitened},
$\widehat{ \mathsf E}$ is traceless and Hermitian and satisfies
$\widehat{\mathsf E}^{\,2}=\mathsf I$.
Thus $ \mathsf W- \mathsf E$ is positive semidefinite of rank one.
Choose a column $\beta$ that is nonzero at the point under
consideration and reduce the rectangle so that it remains nonzero.
Since $\beta$ is a column of $\mathsf W-\mathsf E$, we have
$ \mathsf L\beta=-\beta$. Anticommutation of $\mathsf J$ and $\mathsf L$
then gives $\mathsf L\mathsf J\beta=\mathsf J\beta$.
Thus $\beta$ and $\mathsf J\beta$ are independent, so $\mathsf B$ is invertible.
The Pauli relations give
\begin{equation}
  \mathsf J \mathsf B= \mathsf B\mathsf T,\qquad
  \mathsf L \mathsf B= \mathsf B\tau_2.
 \label{eq:normalization-proof-column-identities}
\end{equation}
The positive Hermitian matrix
$ \mathsf W_B\coloneqq  \mathsf B^{-1} \mathsf W( \mathsf B^\dagger)^{-1}$
commutes with $\mathsf T$ by Eq.~\eqref{eq:general-JW-congruence}.
Hermiticity of
$ \mathsf B^{-1} \mathsf E( \mathsf B^\dagger)^{-1}
=\tau_2 \mathsf W_B$
also makes $ \mathsf W_B$ commute with $\tau_2$.
Hence $ \mathsf W_B=c \mathsf I$ with
$c=1/|\det \mathsf B|$. It follows that $\mathsf X$ defined in
Eq.~\eqref{eq:pd-X-reconstruction}
has unit determinant and reproduces
\begin{equation}
  \mathsf J= \mathsf X\mathsf T \mathsf X^{-1},\qquad
  \mathsf W= \mathsf X \mathsf X^\dagger,\qquad
  \mathsf E= \mathsf X\tau_2 \mathsf X^\dagger.
 \label{eq:direct-auxiliary-matrices}
\end{equation}
This proves the reconstruction formula, with its smooth local
square-root choice.

The frame orientation also gives
$ \mathsf X^{-1} \mathsf B_2( \mathsf X^\dagger)^{-1}=\tau_1$.
Since $\mathsf S_a=\sum_{i=0}^3 \mathsf B_i\mathsf V_{i+1,a}$,
we obtain
\begin{equation}
  \mathsf X^{-1} \mathsf S_a( \mathsf X^\dagger)^{-1}
 = \mathsf R_{0a}(x) \mathsf I+ \mathsf R_{1a}(x)\mathsf T
 + \mathsf Q_{0a}(y)\tau_1+ \mathsf Q_{1a}(y)\tau_2.
 \label{eq:general-separated-sources}
\end{equation}
The coupling coefficients in Eq.~\eqref{eq:general-coupling-target} are
\begin{equation}
 \begin{aligned}
 c_0&=\sum_a d_a(\mathfrak m)\mathsf R_{0a},&
 c_3&=\sum_a d_a(\mathfrak m)\mathsf R_{1a},\\
 c_1&=\sum_a d_a(\mathfrak m)\mathsf Q_{0a},&
 c_2&=\sum_a d_a(\mathfrak m)\mathsf Q_{1a}.
 \end{aligned}
 \label{eq:general-coupling-coefficients}
\end{equation}
\proofend

The matrix $\mathsf J$ defined in
Eq.~\eqref{eq:general-frame-matrices} can also be recovered directly
from any two independent
Hermitian matrices $ \mathsf U, \mathsf V$ in the radial plane:
\begin{equation}
 \begin{gathered}
 d_{UV}^2\coloneqq
 \langle \mathsf U, \mathsf V\rangle^2
 -\det \mathsf U\det \mathsf V>0,\\
  \mathsf J=\pm
 \frac{ \mathsf U\mathsf V^\#
 -\langle \mathsf U, \mathsf V\rangle \mathsf I}{d_{UV}},
 \qquad d_{UV}>0.
 \end{gathered}
 \label{eq:separation-J-reconstruction}
\end{equation}
\equationalias{eq:separation-proof-J-reconstruction}
The sign is chosen to agree with
$\mathsf J=\mathsf B_1\mathsf B_0^{-1}$.
Indeed, the identities
\begin{equation}
  \mathsf U\mathsf V^\#
 + \mathsf V\mathsf U^\#
 =2\langle \mathsf U, \mathsf V\rangle \mathsf I,
 \qquad
 \det( \mathsf U\mathsf V^\#)
 =\det \mathsf U\det \mathsf V
 \label{eq:separation-proof-adjugate-identities}
\end{equation}
and the Cayley--Hamilton identity give
$\operatorname{tr} \mathsf J=0$ and $ \mathsf J^2= \mathsf I$.
Hermiticity gives
$ \mathsf J \mathsf U= \mathsf U \mathsf J^\dagger$ and
$ \mathsf J \mathsf V= \mathsf V \mathsf J^\dagger$.
A transformation $\mathsf M\mapsto\mathsf X\mathsf M\mathsf X^\dagger$
with $\det\mathsf X=1$ taking the positive radial direction
to $ \mathsf I$, followed by a unitary rotation of the other radial
direction to $\mathsf T$, identifies its action on both planes:
\begin{equation}
  \mathsf J \mathsf U_x \mathsf J^\dagger= \mathsf U_x,\qquad
  \mathsf J \mathsf U_y \mathsf J^\dagger=- \mathsf U_y.
 \label{eq:separation-proof-congruence-action}
\end{equation}
Thus the radial plane determines $ \mathsf J$ up to orientation.

To establish necessity, we must show that separation of
$\mathsf F$ implies separation of each source $\mathsf S_a$.
Since the sources are Hermitian,
\begin{equation}
 \begin{aligned}
 \frac{\mathsf F(\mathfrak m)+\mathsf F(\mathfrak m)^\dagger}{2}
 &=\sum_{a=1}^4 \operatorname{Re}d_a(\mathfrak m)\,\mathsf S_a,\\
 \frac{\mathsf F(\mathfrak m)-\mathsf F(\mathfrak m)^\dagger}{2i}
 &=\sum_{a=1}^4 \operatorname{Im}d_a(\mathfrak m)\,\mathsf S_a.
 \end{aligned}
 \label{eq:general-source-recoverability}
\end{equation}
We assume that, by evaluating these equations at suitable mode
labels, we can select four equations with an invertible
coefficient matrix. Solving them recovers the four sources
$\mathsf S_a$. We call this assumption source recoverability.
Since the coefficients $d_a$ are independent of $x,y$,
each source is recovered by a coordinate-independent real
linear combination of these Hermitian matrices.
These operations commute with the transformation
$\mathsf M\mapsto\mathsf X^{-1}\mathsf M(\mathsf X^\dagger)^{-1}$,
so a mode-independent transformation separating $\mathsf F$
also separates every source.
The Gram argument above then gives an admissible decomposition.
Thus, under this recoverability condition and the source basis
assumption, an admissible Gram decomposition is necessary and
sufficient for coupling separation.

The admissible decomposition need not be unique, because the
rank and signature conditions may allow more than one constant
shift in Eq.~\eqref{eq:general-constant-split-freedom}.
For a unique decomposition, the two source planes are fixed,
and the reconstruction above determines $\mathsf J$ up to sign.

\subsection{Transport conditions and determination of \texorpdfstring{$\mathsf A$}{A}}
\label{sec:differential-compatibility}

Retain the assumptions on $\mathscr D$ and $ \mathsf A$ in
Sec.~\ref{sec:general-separation-theorem}.
For a chosen Gram decomposition satisfying the rank and signature
conditions of Sec.~\ref{sec:coupling-construction},
Theorem~\ref{thm:coupling-construction} constructs a smooth
mode-independent matrix $ \mathsf X\in SL(2,\mathbb C)$
that separates the coupling coefficients.
The remaining step is to verify the transport identity
and the separation of the transport function.
To check these two conditions, we determine whether there exist real functions
$\xi(x),\eta(y)$ such that
\begin{equation}
  \mathsf X^{-1}(\mathscr D \mathsf X- \mathsf A \mathsf X)
 =\bigl(\xi(x)+\eta(y)\bigr)\mathsf T.
 \label{eq:general-transport-residual}
\end{equation}
We now determine which Hermitian matrices $ \mathsf A$ can satisfy
this requirement for the chosen coupling separation.
The differential operator and Gram decomposition are held fixed.

The transport identity involves the matrix
\[
  \mathsf H_F\coloneqq (\mathscr D \mathsf X) \mathsf X^{-1}.
\]
For the oriented source frame of
Eq.~\eqref{eq:general-source-frame}, this matrix can be computed
without reconstructing $\mathsf X$:
\begin{equation}
  \mathsf H_F=
 \frac14\sum_{i=0}^3\eta_i
 (\mathscr D \mathsf B_i)\mathsf B_i^\#,
 \qquad (\eta_0,\eta_1,\eta_2,\eta_3)\coloneqq (1,-1,-1,-1).
 \label{eq:source-frame-transport}
\end{equation}
This matrix depends on the chosen separated source frame.
To verify this identity, write $\mathsf B_i=\mathsf X\rho_i\mathsf X^\dagger$ with
$(\rho_0,\rho_1,\rho_2,\rho_3)\coloneqq (\mathsf I,\mathsf T,\tau_1,\tau_2)$.
Since $\rho_i^\#=\eta_i\rho_i$ and $\det \mathsf X=1$, differentiating gives
\[
 \frac14\sum_i\eta_i(\mathscr D\mathsf B_i)\mathsf B_i^\#
 =(\mathscr D\mathsf X)\mathsf X^{-1}
 +\frac14\mathsf X\!\left(\sum_i\rho_i
       (\mathscr D\mathsf X^\dagger)(\mathsf X^\dagger)^{-1}\rho_i\right)\!\mathsf X^{-1}.
\]
The Pauli identity
\[
 \sum_i\rho_i\mathsf M\rho_i=2\operatorname{tr}(\mathsf M)\mathsf I
\]
makes the second term zero: its matrix argument has trace
$\operatorname{tr}\bigl((\mathscr D\mathsf X^\dagger)(\mathsf X^\dagger)^{-1}\bigr)=0$,
as follows by differentiating $\det \mathsf X^\dagger=1$.

To determine the Hermitian coefficient $\mathsf A$ and the real
scalar function $\chi$ in the relation
$\mathsf H_F-\mathsf A=\chi\mathsf J$, we separate its Hermitian
real and imaginary parts.
For any matrix $\mathsf M$, write $\mathsf M_{\mathrm R}\coloneqq (\mathsf M+\mathsf M^\dagger)/2$ and
$\mathsf M_{\mathrm I}\coloneqq (\mathsf M-\mathsf M^\dagger)/(2i)$.
In particular,
\begin{equation}
  \mathsf J= \mathsf J_{\mathrm R}+i \mathsf J_{\mathrm I},
 \qquad
  \mathsf H_F= (\mathsf H_F)_{\mathrm R}+i (\mathsf H_F)_{\mathrm I}.
 \label{eq:hermitian-transport-parts}
\end{equation}
Where $ \mathsf J_{\mathrm I}\ne0$, the imaginary part fixes a
candidate for $\chi$, and the real part then fixes a candidate
for $\mathsf A$:
\begin{equation}
 \chi_F\coloneqq
 \frac{\operatorname{tr}( (\mathsf H_F)_{\mathrm I}
                          \mathsf J_{\mathrm I})}
      {\operatorname{tr}( \mathsf J_{\mathrm I}^2)},
 \qquad
  \mathsf A_F\coloneqq
  (\mathsf H_F)_{\mathrm R}-\chi_F \mathsf J_{\mathrm R}.
 \label{eq:hermitian-transport-candidates}
\end{equation}
The denominator is positive, and $\chi_F$ is real.
Both candidates are determined before testing additive separation.

\begin{theorem}[Determination of the Hermitian coefficient]
\label{thm:hermitian-transport}
Fix the differential operator, an admissible Gram decomposition,
and an oriented separated source frame as above.
Work on an open rectangle where $ \mathsf J_{\mathrm I}\ne0$.
For a smooth mode-independent Hermitian traceless matrix
$ \mathsf A$, the two transport conditions
\[
 \mathsf X^{-1}(\mathscr D\mathsf X-\mathsf A\mathsf X)=\bigl(\xi(x)+\eta(y)\bigr)\mathsf T
\]
hold if and only if
\begin{equation}
  (\mathsf H_F)_{\mathrm I}=\chi_F \mathsf J_{\mathrm I},
 \qquad
  \mathsf A= \mathsf A_F,
 \qquad
 \partial_x\partial_y\chi_F=0.
 \label{eq:hermitian-transport-tests}
\end{equation}
When the first condition holds, $ \mathsf A_F$ and $\chi_F$
are the unique Hermitian matrix and real scalar function
satisfying
$ \mathsf H_F- \mathsf A=\chi \mathsf J$
for this frame, even before additive separation is imposed.
\end{theorem}

\paragraph{Proof.}
The two transport conditions are equivalent to
$ \mathsf H_F- \mathsf A=\chi \mathsf J$ with
$\chi=\xi(x)+\eta(y)$ real.
Taking Hermitian real and imaginary parts gives
\[
  (\mathsf H_F)_{\mathrm R}- \mathsf A
 =\chi \mathsf J_{\mathrm R},
 \qquad
  (\mathsf H_F)_{\mathrm I}=\chi \mathsf J_{\mathrm I}.
\]
Multiplying the second identity by $ \mathsf J_{\mathrm I}$
and taking the trace fixes $\chi=\chi_F$.
The first identity then fixes $ \mathsf A= \mathsf A_F$.
Conversely, the first two conditions in
Eq.~\eqref{eq:hermitian-transport-tests} give
$ \mathsf H_F- \mathsf A=\chi_F \mathsf J$.
The last condition makes the real function $\chi_F$
additively separated on the rectangle.
Tracelessness follows from $\det \mathsf X=1$ and
$\operatorname{tr} \mathsf J=0$.\proofend

If $ \mathsf J_{\mathrm I}=0$ on an open region, Hermiticity
instead requires $ (\mathsf H_F)_{\mathrm I}=0$.
When this holds, compatible matrices have the form
$ \mathsf A= (\mathsf H_F)_{\mathrm R}-\chi \mathsf J$,
with real additively separated $\chi$; uniqueness need not hold.
At zeros of $ \mathsf J_{\mathrm I}$ bordering a non-Hermitian
region, one must use the original two transport conditions and
smoothness, rather than assume that the quotient defining
$\chi_F$ is regular.

\paragraph{Frame freedom for a fixed Gram decomposition.}
\label{app:general-freedom}

We now show that the transformations in
Eq.~\eqref{eq:separation-residual-gauge} exhaust the continuous
freedom for a fixed oriented Gram decomposition, and that the
transport test is independent of this choice.
Fix a Gram decomposition, its time orientation, and the sign of
$ \mathsf J$. Any two unit determinant lifts giving the same
$ \mathsf J$ satisfy
\begin{equation}
  \mathsf Z\coloneqq \mathsf X^{-1} \mathsf X',\qquad
 [ \mathsf Z,\mathsf T]=0,\qquad\det \mathsf Z=1.
 \label{eq:normalization-proof-centralizer}
\end{equation}
In an eigenbasis of $\mathsf T$, the matrix $ \mathsf Z$ is diagonal
with entries $c,c^{-1}$. A local logarithm therefore gives
\begin{equation}
  \mathsf X'= \mathsf X e^{(f+ig)\mathsf T},\qquad f,g\in\mathbb R.
 \label{eq:normalization-proof-local-gauge}
\end{equation}
At this algebraic stage, $f,g$ may depend on both coordinates.
Their action on the frame matrices is
\begin{equation}
 \begin{aligned}
  \mathsf W'&=\cosh(2f) \mathsf W+\sinh(2f) \mathsf J \mathsf W,\\
  \mathsf E'&=\cos(2g) \mathsf E+i\sin(2g) \mathsf J \mathsf E.
 \end{aligned}
 \label{eq:normalization-proof-frame-freedom}
\end{equation}
For two separated source frames, the radial coefficient vectors
depend only on $x$ and span the radial plane, while the angular
coefficient vectors depend only on $y$ and span the angular plane.
Differentiating the boost and rotation with respect to the opposite
coordinate therefore gives $\partial_y f=0$ and $\partial_x g=0$:
each generator is invertible on its two-dimensional plane.
Equivalently, the one-variable Gram factors differ by a radial boost
and an angular rotation. Thus the remaining continuous freedom is
precisely Eq.~\eqref{eq:separation-residual-gauge}.

Conversely, this freedom allows us to choose $\mathsf W$ and
$\mathsf E$ directly, without first specifying the Gram factors.
Choose smooth mode-independent Hermitian matrices in the radial
and angular planes, respectively, such that
\begin{equation}
 \mathsf W>0,\qquad\det\mathsf W=1,\qquad\det\mathsf E=-1.
 \label{eq:general-direct-frame-normalization}
\end{equation}
To obtain a separated source frame, require in addition that
\begin{equation}
 \partial_y\langle\mathsf W,\mathsf S_a\rangle=0,\qquad
 \partial_x\langle\mathsf E,\mathsf S_a\rangle=0
 \quad(a=1,\ldots,4).
 \label{eq:general-direct-frame-separation}
\end{equation}
Relative to any separated frame constructed in
Theorem~\ref{thm:coupling-construction}, these choices are related
by the boost and rotation above. The additional conditions force
their parameters to depend only on $x$ and $y$, respectively.
Completing the frame as
$(\mathsf W,\mathsf J\mathsf W,i\mathsf J\mathsf E,\mathsf E)$
therefore gives another separated frame.
The corresponding factors in Eq.~\eqref{eq:general-gram-factors}
are recovered from the source coefficients as
\begin{align}
 R_{0a}&=\langle\mathsf W,\mathsf S_a\rangle,
 &R_{1a}&=-\langle\mathsf J\mathsf W,\mathsf S_a\rangle,
 \label{eq:pd-radial-frame-coefficients}\\
 Q_{0a}&=-\langle i\mathsf J\mathsf E,\mathsf S_a\rangle,
 &Q_{1a}&=-\langle\mathsf E,\mathsf S_a\rangle.
 \label{eq:pd-angular-frame-coefficients}
\end{align}
These factors reproduce the specified Gram decomposition,
and the construction in Eq.~\eqref{eq:general-source-frame}
recovers the chosen frame matrices.

Once the oriented frame matrices $(\mathsf J,\mathsf W,\mathsf E)$
are fixed, $\mathsf X$ is determined up to sign.
Indeed, the ratio of two lifts commutes with $\mathsf T$,
is unitary, and preserves the chosen transverse Pauli direction;
it is therefore $\pm\mathsf I$.

To examine the transport conditions under this change of frame,
set $\delta\coloneqq \partial_x f+\partial_y g$.
Then $ \mathsf J'= \mathsf J$ and
$ \mathsf H_F'= \mathsf H_F+\delta \mathsf J$.
Since $\delta$ is real and additively separated, the relation
$ \mathsf H_F- \mathsf A=\chi \mathsf J$ is carried to
$ \mathsf H_F'- \mathsf A=(\chi+\delta) \mathsf J$.
Thus both transport conditions are invariant under these changes.
On the non-Hermitian region, the candidates transform as
\begin{equation}
 \chi_F'=\chi_F+\delta,\qquad
  \mathsf A_F'= \mathsf A_F.
 \label{eq:hermitian-transport-gauge}
\end{equation}
The mismatch
$ (\mathsf H_F)_{\mathrm I}-\chi_F \mathsf J_{\mathrm I}$
and the mixed derivative of $\chi_F$ are also unchanged.
Hence the three tests in
Eq.~\eqref{eq:hermitian-transport-tests} depend on the chosen
Gram decomposition, but not on its separated frame.
The compatible matrix $ \mathsf A_F$ is fixed by that decomposition
and $\mathscr D$, whereas $\chi_F$ depends on the representative.

\label{app:separation-scope}
All statements above are local. The source construction and the
necessity and uniqueness arguments require a nonsingular source Gram
matrix; the separation theorem does not.
Degeneracy of the chosen sources does not itself prove failure
of separation; another presentation or an independently justified
extension may be available. These theorems do not assert global
single-valuedness, regularity in a spacetime frame, or physical boundary
conditions. Nor do they prove spectral completeness.
For general complex source coefficients $d_a(\mathfrak m)$, real mode
parameters need not give $c_1+i c_2=-\overline{(c_1-i c_2)}$, the condition required
for formal symmetry of $\mathcal L$ with real $\eta$. Self-adjointness
and spectral properties require a specified operator domain as well.

\Needspace{24\baselineskip}
\subsection{The complete criterion and its use}
\label{sec:complete-criterion}
The preceding results separate the construction into an algebraic
problem for the sources and a transport test involving $\mathsf A$.
They combine into the following criterion.

\begin{theorem}[Constructive separation criterion]
\label{thm:complete-criterion}
Consider the chiral system with the source basis and recoverability
assumptions of Sec.~\ref{sec:coupling-construction}.
Fix the constant Hermitian axis $\mathsf T$.
A smooth, mode-independent matrix $\mathsf X\in SL(2,\mathbb C)$
satisfying the three conditions of
Theorem~\ref{thm:chiral-separation} exists locally if and only if
the following two conditions hold on an open rectangle.
\begin{enumerate}
\item
The source Gram matrix admits a real symmetric decomposition
\begin{equation}
 \Gamma(x,y)=\Gamma_x(x)+\Gamma_y(y),
 \qquad
 \operatorname{rank}\Gamma_x=\operatorname{rank}\Gamma_y=2,
 \label{eq:complete-gram-condition}
\end{equation}
with nonzero signatures $(+,-)$ and $(-,-)$, respectively.

\item
For a corresponding oriented separated source frame, there
exists a smooth real, mode-inde\-pendent function $\chi$ such that
\begin{equation}
 (\mathsf H_F)_{\mathrm I}=\chi\mathsf J_{\mathrm I},
 \qquad
 \mathsf A=(\mathsf H_F)_{\mathrm R}-\chi\mathsf J_{\mathrm R},
 \qquad
 \partial_x\partial_y\chi=0.
 \label{eq:complete-transport-condition}
\end{equation}
Here $\mathsf J$ and $\mathsf H_F$ are computed from the source
frame using Eqs.~\eqref{eq:general-frame-matrices} and
\eqref{eq:source-frame-transport}; the subscripts denote the
Hermitian real and imaginary parts used in
Sec.~\ref{sec:differential-compatibility}.
\end{enumerate}
For a fixed oriented decomposition, the transport test has the
same outcome for every separated frame.
Without source recoverability, these conditions remain sufficient.
\end{theorem}

\paragraph{Proof.}
Theorem~\ref{thm:coupling-construction} constructs a matrix
$\mathsf X$ that separates the coupling.
The second condition is equivalent to
$\mathsf H_F-\mathsf A=\chi\mathsf J$ with
$\chi=\xi(x)+\eta(y)$ real.
Since $\mathsf H_F=(\mathscr D\mathsf X)\mathsf X^{-1}$
and $\mathsf J=\mathsf X\mathsf T\mathsf X^{-1}$,
this gives the transport identity and separation of its
transport function. Theorem~\ref{thm:chiral-separation}
then gives the separated equations.
Conversely, the necessity argument in
Sec.~\ref{sec:coupling-construction} gives the Gram decomposition,
and the transport identity gives the second condition.
Independence of the separated frame was established in
Sec.~\ref{sec:differential-compatibility}.\proofend

For a specified pair $(\mathsf F,\mathsf A)$, the criterion
can be applied as follows.
\begin{enumerate}
\item
Compute $\Gamma$ and check $\partial_x\partial_y\Gamma=0$.
\item
Determine the constant shifts in
Eq.~\eqref{eq:general-constant-split-freedom} that give
 the required ranks and signatures.
\item
For each admissible decomposition, construct an oriented
separated source frame, compute $\mathsf J$ and $\mathsf H_F$,
and test Eq.~\eqref{eq:complete-transport-condition}.
Where $\mathsf J_{\mathrm I}\ne0$,
Theorem~\ref{thm:hermitian-transport} gives the explicit test;
the remaining cases are discussed in
Sec.~\ref{sec:differential-compatibility}.
\item
If the test succeeds, construct $\mathsf X$ and use the freedom
in Eq.~\eqref{eq:separation-residual-gauge} to simplify its
entries or the separated coefficients.
\end{enumerate}
The transport test does not require reconstructing or
differentiating $\mathsf X$.
Failure for one admissible decomposition does not exclude
success for another. Under source recoverability, failure for
every admissible decomposition rules out separation within
the specified transformation class.
For lower-rank source families, the scope of
Theorem~\ref{thm:chiral-separation} is discussed at the end of
Sec.~\ref{sec:differential-compatibility}.

\section{Pleba\'nski--Demia\'nski background and radiative system}
\label{sec:background}

To apply the general criterion developed in
Sec.~\ref{sec:separation-structure}, we first need to express
the PD radiative equations in the chiral form
\eqref{eq:general-chiral-system}.
Our starting point is the local Newman--Penrose reduction used
for Kerr--Newman--de Sitter perturbations in Sec.~2.3 of Ref.~\cite{Yoshida:2026yly}.
Its application to Pleba\'nski--Demia\'nski requires the background coefficients and
the rescaling of the radiative fields to be specified anew.
We first collect the geometric data, then give the coordinate
equations after rescaling the fields by powers of the conformal factor,
and finally identify the matrices needed for the separation problem.

We adopt the curvature and Newman--Penrose conventions of
Ref.~\cite{Yoshida:2026yly}, with signature $(-+++)$ and
$c=G=k_{\mathrm e}=1$, where $k_{\mathrm e}$ is Coulomb's constant.
The equations to be linearized are the source-free
Einstein--Maxwell equations
\begin{equation}
 \begin{gathered}
 R_{\mu\nu}-\frac12\mathcal R_{\rm sc}g_{\mu\nu}+\Lambda g_{\mu\nu}
 =2\left(F_{\mu\rho}F_\nu{}^\rho
 -\frac14g_{\mu\nu}F_{\rho\sigma}F^{\rho\sigma}\right),\\
 \nabla_\mu F^{\mu\nu}=0,\qquad \bm d\bm F=0.
 \end{gathered}
 \label{eq:bg-field-equations}
\end{equation}
Here $\mathcal R_{\rm sc}$ is the Ricci scalar. The cosmological constant is held
fixed under linearization, so the relation $\mathcal R_{\rm sc}=4\Lambda$ also
fixes the scalar-curvature perturbation.

\subsection{Pleba\'nski--Demia\'nski geometry and principal frame}
\label{sec:geometry}

In the Podolsk\'y--Vr\'atn\'y coordinates
$(t,r,\theta,\phi)$~\cite{Podolsky:2022xxd}, the rotating
Pleba\'nski--Demia\'nski metric and electromagnetic potential are
\begin{align}
 \bm g={}&\frac1{\Omega^2}\biggl\{
 -\frac{\mathcal{Q}}{\Sigma}
 \Bigl(\bm dt-\bigl(a\sin^2\theta+2\ell(1-\cos\theta)\bigr)\bm d\phi\Bigr)^2
 +\frac\Sigma{\mathcal{Q}}(\bm dr)^2
 \nonumber\\
 &\hspace{25mm}+\frac\Sigma{\mathcal P}(\bm d\theta)^2
 +\frac{\mathcal P\sin^2\theta}{\Sigma}
 \Bigl(a\bm dt-\bigl(r^2+(a+\ell)^2\bigr)\bm d\phi\Bigr)^2\biggr\},
 \label{eq:bg-metric}\\
 \bm A={}&-\frac{{Q_e}r-{Q_m}\ell}{\Sigma}
 \Bigl(\bm dt-\bigl(a\sin^2\theta+2\ell(1-\cos\theta)\bigr)\bm d\phi\Bigr)
 \nonumber\\
 &+\frac{{Q_m}\cos\theta}{\Sigma}
 \Bigl(a\bm dt-\bigl(r^2+(a+\ell)^2\bigr)\bm d\phi\Bigr),
 \qquad \bm F\coloneqq \bm d\bm A.
 \label{eq:bg-potential}
\end{align}
Here $M$ is the mass, $a$ the rotation parameter, $\ell$ the NUT
parameter, $\alpha$ the acceleration, ${Q_e}$ the electric charge,
${Q_m}$ the magnetic charge, and $\Lambda$ the cosmological constant.
The potential includes the correction in
Ref.~\cite{Podolsky:2023erratum}, with its magnetic parameter replaced
by $-{Q_m}$ to implement the convention $\Phi_1\propto {Q_e}+i{Q_m}$.
The conformal factor and the function $\Sigma$ are
\begin{equation}
 \Omega\coloneqq1-brp,\qquad \Sigma\coloneqq r^2+p^2.
 \label{eq:bg-notation}
\end{equation}
The radial and angular metric functions are
\begin{align}
 \mathcal{Q}(r)\coloneqq{}&
 (r^2-2Mr+a^2-\ell^2+{Q}^2)
 \bigl(1+b(a-\ell)r\bigr)\bigl(1-b(a+\ell)r\bigr)
 \nonumber\\
 &-\frac{\Lambda}{3}r^2
 \bigl(r^2+2b\ell(a^2-\ell^2)r+a^2+3\ell^2\bigr),
 \label{eq:bg-radial-function}\\
 \mathcal P(\theta)\coloneqq{}&
 1-2\left(bM-\frac{\Lambda\ell}{3}\right)p
 +\left(b^2(a^2-\ell^2+{Q}^2)+\frac\Lambda3\right)p^2.
 \label{eq:bg-angular-function}
\end{align}
These expressions use the combinations
\begin{equation}
 {Q}\coloneqq\sqrt{{Q_e}^2+{Q_m}^2},\qquad
 b\coloneqq\frac{\alpha a}{a^2+\ell^2},\qquad
 p(\theta)\coloneqq\ell+a\cos\theta,
 \label{eq:bg-parameters}
\end{equation}
where $p$ is a function of $\theta$, not an independent coordinate.

All calculations are on a local gauge and coordinate patch
satisfying
\begin{equation}
 a>0,\qquad {Q}>0,\qquad 0<\theta<\pi,\qquad
 \mathcal{Q}>0,\quad\mathcal P>0,\quad\Omega>0,\quad\Sigma>0.
 \label{eq:bg-local-domain}
\end{equation}
These inequalities select positive real square roots and keep
the tetrad below regular. The condition ${Q}>0$ will also be
needed to fix the perturbed tetrad.
The magnetic potential is understood on this gauge patch;
the time and azimuthal identifications are specified in
Sec.~\ref{sec:global-angular}. Extensions through horizons or
$\Omega=0$ are not needed for the local construction.
No particular horizon structure is assumed.

\paragraph{Principal frame.}
A principal null tetrad with an affinely parametrized
$\bm k_{\mathrm{NP}}$ is
\begin{align}
 \bm k_{\mathrm{NP}}\coloneqq {}&\Omega^2
 \left(\frac{r^2+(a+\ell)^2}{\mathcal{Q}}\bm\partial_t
 +\bm\partial_r+\frac a{\mathcal{Q}}\bm\partial_\phi\right),
 \label{eq:bg-tetrad-k}\\
 \bm l_{\mathrm{NP}}\coloneqq {}&
 \frac{\bigl(r^2+(a+\ell)^2\bigr)\bm\partial_t
 -\mathcal{Q}\bm\partial_r+a\bm\partial_\phi}{2\Sigma},
 \label{eq:bg-tetrad-l}\\
 \bm m_{\mathrm{NP}}\coloneqq {}&\frac\Omega{\sqrt2z}
 \left(\frac{i\bigl(a\sin^2\theta+2\ell(1-\cos\theta)\bigr)}{\sqrt{\mathcal P}\sin\theta}
 \bm\partial_t+\sqrt{\mathcal P}\bm\partial_\theta
 +\frac{i}{\sqrt{\mathcal P}\sin\theta}\bm\partial_\phi\right).
 \label{eq:bg-tetrad-m}
\end{align}
Here
\begin{equation}
 z\coloneqq r+ip.
 \label{eq:bg-square-roots}
\end{equation}
The tetrad obeys $\bm k_{\mathrm{NP}}\cdot\bm l_{\mathrm{NP}}=-1$
and $\bm m_{\mathrm{NP}}\cdot\overline{\bm m}_{\mathrm{NP}}=1$,
with all other inner products vanishing.

We use $\Psi_i$ for Weyl scalars, $\Phi_i$ for Maxwell scalars,
and $\Phi_{ij}$ for the scalars of the traceless Ricci tensor.
Their definitions and those of the spin coefficients with
subscript $\mathrm{NP}$ are given in Sec.~2.2 of
Ref.~\cite{Yoshida:2026yly}, following Ref.~\cite{Stephani:2003}.
In particular, frame subscripts refer to
$(\bm k_{\mathrm{NP}},\bm l_{\mathrm{NP}},\bm m_{\mathrm{NP}},
\overline{\bm m}_{\mathrm{NP}})$, and
$\Phi_1\coloneqq (F_{kl}+F_{\bar m m})/2$.
The Einstein--Maxwell relation is $\Phi_{ij}=2\Phi_i\bar\Phi_j$.

\paragraph{Background coefficients.}
Projection onto this tetrad gives
\begin{equation}
 \Phi_1=\frac{({Q_e}+i{Q_m})\Omega^2}{2\bar z^2},\qquad
 \Psi_2=-\frac{\Omega^3}{\bar z^3}
 \left(\mathcal M-\frac{{Q}^2(1+brp)}z\right),\qquad
 \Phi_{11}=\frac{{Q}^2\Omega^4}{2\Sigma^2}.
 \label{eq:bg-curvature-scalars}
\end{equation}
Here $\bar z=r-ip$ and $\Sigma=z\bar z$.
The complex coefficient
\begin{equation}
 \mathcal M\coloneqq
 M+b\ell(a^2-\ell^2+{Q}^2)
 -i\left(\ell-bM(a^2-\ell^2)
 +\frac{\Lambda\ell}{3}(a^2-\ell^2)\right)
 \label{eq:bg-complex-mass}
\end{equation}
is determined entirely by the metric parameters.
All other Weyl, Maxwell, and traceless Ricci scalars vanish.
The principal directions are therefore aligned with the
background Maxwell field.
The charge factor ${Q_e}+i{Q_m}$ fixes our magnetic convention consistently
with the potential and tetrad. In particular
$2\Phi_1\bar\Phi_1={Q}^2\Omega^4/(2\Sigma^2)=\Phi_{11}$;
the metric, curvature scalars and chiral matrices depend on the charges
only through ${Q}^2={Q_e}^2+{Q_m}^2$.
The NUT parameter allows $\Lambda$ to occur explicitly in
$\mathcal M$ and hence in $\Psi_2$.

The spin coefficients needed to evaluate the radiative equations are
\begin{equation}
 \begin{gathered}
 \rho_{\mathrm{NP}}=-\frac{\Omega(1-ibp^2)}{\bar z},\qquad
 \mu_{\mathrm{NP}}=-\frac{\mathcal{Q}(1-ibp^2)}{2\Sigma \bar z\Omega},\qquad
 \gamma_{\mathrm{NP}}=\mu_{\mathrm{NP}}
 +\frac{\partial_r \mathcal{Q}}{4\Sigma}+\frac{bp\mathcal{Q}}{\Sigma\Omega},\\[3pt]
 \beta_{\mathrm{NP}}=\frac{\Omega(\partial_\theta \sqrt{\mathcal P}+\sqrt{\mathcal P}\cot\theta)}{2\sqrt2z},
 \qquad\pi_{\mathrm{NP}}=\frac{i\sqrt{P}(1+ibr^2)}{\sqrt2\bar z^2},\qquad
 \tau_{\mathrm{NP}}=-\frac{i\sqrt{P}(1+ibr^2)}{\sqrt2\Sigma},\\[3pt]
 \alpha_{\mathrm{NP}}=\pi_{\mathrm{NP}}-\bar\beta_{\mathrm{NP}}
 +\frac{\sqrt2br\sqrt{P}}{\bar z},\qquad
 \kappa_{\mathrm{NP}}=\sigma_{\mathrm{NP}}=\lambda_{\mathrm{NP}}
 =\nu_{\mathrm{NP}}=\epsilon_{\mathrm{NP}}=0.
 \end{gathered}
 \label{eq:np-background-coefficients}
\end{equation}
Here
\begin{equation}
 P(\theta)\coloneqq a^2\sin^2\theta\,\mathcal P(\theta),\qquad
 \sqrt{P(\theta)}=a\sin\theta\sqrt{\mathcal P(\theta)}.
 \label{eq:bg-angular-P}
\end{equation}
These are the connection coefficients to be used in the
radiative equations, including their acceleration corrections.

\subsection{Radiative variables and conformal rescaling}
\label{sec:radiative}

Using the background coefficients obtained in the preceding
subsection, we now write the coupled radiative equations
in coordinate form.
We rescale the perturbation variables by powers of the conformal
factor to cancel the additional derivative terms it introduces.
The reduction of Sec.~2.3 of Ref.~\cite{Yoshida:2026yly}
uses the Einstein--Maxwell identities in a principal frame
with vanishing background radiative scalars,
$\kappa_{\mathrm{NP}}=\sigma_{\mathrm{NP}}
=\lambda_{\mathrm{NP}}=\nu_{\mathrm{NP}}=0$, and
$\Phi_1\ne0$.
The background values above meet these conditions.
We can therefore use its phantom-gauge construction
(see also Ref.~\cite{Chandrasekhar:1983}) to impose
\begin{equation}
 \Phi_0^{(1)}=\Phi_2^{(1)}=0.
 \label{eq:bg-phantom}
\end{equation}
The superscript $(1)$ denotes a linear perturbation.
The required infinitesimal null rotations are well defined
because ${Q}>0$ makes the background $\Phi_1$ nonzero.
Its complex phase presents no obstruction, so the same choice
is available with magnetic charge.
Equation~\eqref{eq:bg-phantom} fixes the null rotation freedom
of the perturbed tetrad at linear order;
it does not remove the electromagnetic perturbation.

We take Eqs.~(2.28)--(2.30) and (2.36) of
Ref.~\cite{Yoshida:2026yly} as the starting Newman--Penrose equations:
two linearized Bianchi identities,
one linearized Ricci identity, and the fourth relation obtained
from Maxwell compatibility and the Ricci identities.
Their derivation uses the conditions just listed and does not
require the explicit Kerr--Newman--de Sitter metric functions.
In particular, it introduces no division by
$3\Psi_2\pm2\Phi_{11}$ and no independent scalar-curvature term.
The dependence on $\Lambda$ is retained through the Pleba\'nski--Demia\'nski
background coefficients.
These equations close on four radiative variables in each
spin sector and provide a necessary subsystem of the
linearized Einstein--Maxwell equations; reconstruction of
all metric and electromagnetic perturbations is not implied.

The positive-spin variables are
$(\Psi_0^{(1)},\Psi_1^{(1)},\kappa_{\mathrm{NP}}^{(1)},
\sigma_{\mathrm{NP}}^{(1)})$, with weights $(2,1,1,2)$.
A weight $s$ means multiplication by $e^{is\vartheta}$ when
$\bm m_{\mathrm{NP}}$ is replaced by
$e^{i\vartheta}\bm m_{\mathrm{NP}}$ with the real null vectors
fixed.
The negative-spin system follows from the tetrad exchange
\begin{equation}
 (\bm k_{\mathrm{NP}},\bm l_{\mathrm{NP}},
 \bm m_{\mathrm{NP}},\overline{\bm m}_{\mathrm{NP}})'
 \coloneqq (\bm l_{\mathrm{NP}},\bm k_{\mathrm{NP}},
 \overline{\bm m}_{\mathrm{NP}},\bm m_{\mathrm{NP}}).
 \label{eq:np-prime}
\end{equation}
This is the exchange used in Sec.~2.3 of
Ref.~\cite{Yoshida:2026yly}. It is performed in the Newman--Penrose equations
before imposing $\epsilon_{\mathrm{NP}}=0$; it includes
$\epsilon_{\mathrm{NP}}'=-\gamma_{\mathrm{NP}}$.
We now specify the coordinate operators and rescaled fields
for both systems.

\paragraph{Mode operators.}
For fields with Fourier dependence $e^{-i\omega t+im\phi}$,
the background directional derivatives along
$(\bm k_{\mathrm{NP}},\bm l_{\mathrm{NP}},\bm m_{\mathrm{NP}},
\overline{\bm m}_{\mathrm{NP}})$ take the form
\begin{equation}
 D=\Omega^2\mathcal D_0,\qquad
 \Delta=-\frac{\mathcal{Q}}{2\Sigma}\mathcal D_0^\dagger,\qquad
 \delta=\frac\Omega{\sqrt2z}\mathcal L_0^\dagger,\qquad
 \bar\delta=\frac\Omega{\sqrt2\bar z}\mathcal L_0.
 \label{eq:bg-directional-derivatives}
\end{equation}
Here the radial and angular operators are defined, for integer $s$, by
\begin{align}
 \mathcal D_s&\coloneqq\partial_r+\frac{iK}{\mathcal{Q}}
 +s\frac{\partial_r \mathcal{Q}}{\mathcal{Q}},&
 \mathcal D_s^\dagger&\coloneqq\partial_r-\frac{iK}{\mathcal{Q}}
 +s\frac{\partial_r \mathcal{Q}}{\mathcal{Q}},
 \label{eq:bg-radial-operators}\\
 \mathcal L_s&\coloneqq \sqrt{\mathcal P}\partial_\theta+W
 +s(\partial_\theta \sqrt{\mathcal P}+\sqrt{\mathcal P}\cot\theta),&
 \mathcal L_s^\dagger&\coloneqq \sqrt{\mathcal P}\partial_\theta-W
 +s(\partial_\theta \sqrt{\mathcal P}+\sqrt{\mathcal P}\cot\theta).
 \label{eq:bg-angular-operators}
\end{align}
The mode-dependent functions in these operators are
\begin{equation}
 K(r)\coloneqq-\bigl(r^2+(a+\ell)^2\bigr)\omega+am,\qquad
 W(\theta)\coloneqq
 \frac{m-\bigl(a\sin^2\theta+2\ell(1-\cos\theta)\bigr)\omega}{\sqrt{\mathcal P}\sin\theta}.
 \label{eq:bg-mode-potentials}
\end{equation}

The dagger reverses the displayed mode term at fixed
$\omega$; it denotes neither an adjoint nor complex conjugation.
The local frequency $\omega$ may be complex. The global restrictions
on $\omega,m$ are specified in Sec.~\ref{sec:global-angular}.

\paragraph{Conformal weights and coordinate equations.}
Acceleration introduces derivatives of $\Omega$ into the
Newman--Penrose equations. They can be absorbed by assigning different
powers of $\Omega$ to the four fields in each spin sector.
For positive spin weights we use
\begin{equation}
 \begin{aligned}
 \widetilde\Psi_0&\coloneqq\Omega^{-5}\Psi_0^{(1)},&
 \widetilde\Psi_1&\coloneqq\sqrt2\bar z\Omega^{-4}\Psi_1^{(1)},\\
 \widetilde\kappa&\coloneqq
 \frac{\kappa_{\mathrm{NP}}^{(1)}}{\sqrt2\bar z^2\Omega^3},&
 \widetilde\sigma&\coloneqq
 \frac{z\sigma_{\mathrm{NP}}^{(1)}}{\bar z^2\Omega^2}.
 \end{aligned}
 \label{eq:bg-positive-fields}
\end{equation}
For negative spin weights we use
\begin{equation}
 \begin{aligned}
 \widetilde\Psi_4&\coloneqq \bar z^4\Omega^{-1}\Psi_4^{(1)},&
 \widetilde\Psi_3&\coloneqq\frac{\bar z^3\Omega^{-2}}{\sqrt2}\Psi_3^{(1)},\\
 \widetilde\nu&\coloneqq\frac{\Omega z\bar z}{\sqrt2}\nu_{\mathrm{NP}}^{(1)},&
 \widetilde\lambda&\coloneqq\frac{\bar z}{2}\lambda_{\mathrm{NP}}^{(1)}.
 \end{aligned}
 \label{eq:bg-negative-fields}
\end{equation}
The latter fields have weights $(-2,-1,-1,-2)$.
These factors extend the field rescalings of
Ref.~\cite{Yoshida:2026yly} to the accelerated background.

The curvature combinations entering the radiative equations can
be written as
\begin{equation}
 \frac{\bar z^3}{\Omega^3}(3\Psi_2-2\Phi_{11})=-\mathcal F_+,
 \qquad
 \frac{\bar z^3}{\Omega^3}(3\Psi_2+2\Phi_{11})=-\mathcal F_-,
 \label{eq:bg-F-identities}
\end{equation}
where
\begin{equation}
 \mathcal F_\pm\coloneqq
 3\left(\mathcal M-\frac{{Q}^2(1+brp)}z\right)
 \pm\frac{{Q}^2\Omega \bar z}{z^2}.
 \label{eq:bg-Fpm}
\end{equation}

Substituting the Pleba\'nski--Demia\'nski background coefficients
and the field rescalings in Eq.~\eqref{eq:bg-positive-fields}
into Eqs.~(2.28)--(2.30) and (2.36) of
Ref.~\cite{Yoshida:2026yly}, the terms involving derivatives
of $\Omega$ cancel, yielding
\begin{subequations}\label{eq:bg-positive-system}
\begin{align}
 (\mathcal L_2-3i\sqrt{P}/\bar z)\widetilde\Psi_0
 -(\mathcal D_0+3/\bar z)\widetilde\Psi_1
 &=-2\mathcal F_+\widetilde\kappa,
 \label{eq:bg-positive-a}\\
 \mathcal{Q}(\mathcal D_2^\dagger-3/\bar z)\widetilde\Psi_0
 +(\mathcal L_{-1}^\dagger+3i\sqrt{P}/\bar z)\widetilde\Psi_1
 &=2\mathcal F_-\widetilde\sigma,
 \label{eq:bg-positive-b}\\
 (\mathcal D_0+3/\bar z)\widetilde\sigma
 -(\mathcal L_{-1}^\dagger+3i\sqrt{P}/\bar z)\widetilde\kappa
 &=\frac{\Omega z}{\bar z^2}\widetilde\Psi_0,
 \label{eq:bg-positive-c}\\
 \mathcal{Q}(\mathcal D_2^\dagger-3/\bar z)\widetilde\kappa
 +(\mathcal L_2-3i\sqrt{P}/\bar z)\widetilde\sigma
 &=\frac{2\Omega z}{\bar z^2}\widetilde\Psi_1.
 \label{eq:bg-positive-d}
\end{align}
\end{subequations}

The corresponding negative-spin equations are
\begin{subequations}\label{eq:bg-negative-system}
\begin{align}
 (\mathcal L_2^\dagger-3i\sqrt{P}/\bar z)\widetilde\Psi_4
 +\mathcal{Q}(\mathcal D_{-1}^\dagger+3/\bar z)\widetilde\Psi_3
 &=2\mathcal F_+\widetilde\nu,
 \label{eq:bg-negative-a}\\
 (\mathcal D_0-3/\bar z)\widetilde\Psi_4
 -(\mathcal L_{-1}+3i\sqrt{P}/\bar z)\widetilde\Psi_3
 &=2\mathcal F_-\widetilde\lambda,
 \label{eq:bg-negative-b}\\
 \mathcal{Q}(\mathcal D_{-1}^\dagger+3/\bar z)\widetilde\lambda
 +(\mathcal L_{-1}+3i\sqrt{P}/\bar z)\widetilde\nu
 &=\frac{\Omega z}{\bar z^2}\widetilde\Psi_4,
 \label{eq:bg-negative-c}\\
 (\mathcal D_0-3/\bar z)\widetilde\nu
 -(\mathcal L_2^\dagger-3i\sqrt{P}/\bar z)\widetilde\lambda
 &=\frac{2\Omega z}{\bar z^2}\widetilde\Psi_3.
 \label{eq:bg-negative-d}
\end{align}
\end{subequations}

The $\Omega$ weights remove the additional derivative terms,
while acceleration remains in the metric functions and in
the algebraic factors $\Omega$ and $\mathcal F_\pm$.
Thus the derivative combinations have the Kerr--Newman--de Sitter form,
with the background dependence displayed explicitly above.

\subsection{Chiral system on the Pleba\'nski--Demia\'nski background}
\label{sec:chiral-system}

To formulate the matrix separation problem, we use the chiral
arrangement introduced by Hintz~\cite{Hintz:2026jiw} and adopted
in Sec.~2.4 of Ref.~\cite{Yoshida:2026yly}.
For either spin sector, the coordinate equations,
Eqs.~\eqref{eq:bg-positive-system} or
\eqref{eq:bg-negative-system}, take the common form
\begin{equation}
 (\mathscr D- \mathsf A)y_+^{(\varsigma)}
 = \mathsf F^{(\varsigma)}y_-^{(\varsigma)},\qquad
 (\overline{\mathscr D}+ \mathsf A)y_-^{(\varsigma)}
 =\bigl(\mathsf F^{(\varsigma)}\bigr)^\#y_+^{(\varsigma)}.
 \label{eq:bg-chiral-system}
\end{equation}
Here the chiral variables are the following combinations of the
conformally rescaled fields:
\begin{align}
 y_\pm^{(+)}&\coloneqq \mathcal{Q}^{5/4}\mathcal P^{1/4}\sqrt{\sin\theta}
 \begin{pmatrix}
 \widetilde\Psi_0\pm i\widetilde\Psi_1/{\sqrt{\mathcal Q}}\\
 -2{Q}(\widetilde\kappa\pm i\widetilde\sigma/{\sqrt{\mathcal Q}})
 \end{pmatrix},
 \label{eq:bg-chiral-plus}\\
 y_\pm^{(-)}&\coloneqq \mathcal{Q}^{-3/4}\mathcal P^{1/4}\sqrt{\sin\theta}
 \begin{pmatrix}
 \widetilde\Psi_4\mp i{\sqrt{\mathcal Q}}\widetilde\Psi_3\\
 2{Q}(\widetilde\nu\mp i{\sqrt{\mathcal Q}}\widetilde\lambda)
 \end{pmatrix}.
 \label{eq:bg-chiral-minus}
\end{align}
The superscript $\varsigma=\pm1$ labels the sign of the original
spin weights. The subscripts $+$ and $-$ correspond to the
principal differential operators
\begin{equation}
 \mathscr D\coloneqq {\sqrt{\mathcal Q}}\partial_r-i\sqrt{\mathcal P}\partial_\theta,\qquad
 \overline{\mathscr D}\coloneqq {\sqrt{\mathcal Q}}\partial_r+i\sqrt{\mathcal P}\partial_\theta.
 \label{eq:bg-chiral-derivatives}
\end{equation}
The normalized coordinates of
Sec.~\ref{sec:general-separation-theorem} are locally defined by
\[
 dx=\frac{dr}{\sqrt{\mathcal Q(r)}},\qquad
 dy=\frac{d\theta}{\sqrt{\mathcal P(\theta)}}.
\]
Thus $\mathscr D=\partial_x-i\partial_y$, and separate
dependence on $x,y$ is equivalent to separate dependence
on $r,\theta$.

The scalar factors in Eqs.~\eqref{eq:bg-chiral-plus} and
\eqref{eq:bg-chiral-minus} account for derivatives of
$\mathcal{Q}$ and $\sqrt{\mathcal P}\sin\theta$.
The use of ${Q}=\sqrt{{Q_e}^2+{Q_m}^2}>0$ in the second component fixes
its constant normalization; no duality rotation of the
background Maxwell field is involved.

The Pleba\'nski--Demia\'nski coefficients in this system are
\begin{equation}
  \mathsf A=\begin{pmatrix}0&-i\mathcal V\\i\overline{\mathcal V}&0\end{pmatrix},
 \qquad \mathcal V\coloneqq \frac{{Q}\Omega\bar z}{z^2},
 \label{eq:bg-chiral-A}
\end{equation}
and
\begin{equation}
  \mathsf F^{(\varsigma)}\coloneqq c^{(\varsigma)} \mathsf I
 - \mathsf B^{(\varsigma)},\qquad
  \mathsf B^{(\varsigma)}\coloneqq i
 \begin{pmatrix}
 \beta^{(\varsigma)}&\displaystyle\frac3{Q}
 \left(\mathcal M-\frac{{Q}^2(1+brp)}z\right)\\[4pt]
 \displaystyle\frac{3{Q}\Omega z}{\bar z^2}&-\beta^{(\varsigma)}
 \end{pmatrix},
 \label{eq:bg-chiral-F}
\end{equation}
with $ \mathsf I$ the identity matrix and
\begin{equation}
 \begin{aligned}
 \beta^{(\varsigma)}&\coloneqq-\varsigma W
 -\frac32(\partial_\theta \sqrt{\mathcal P}+\sqrt{\mathcal P}\cot\theta)+\frac{3i\sqrt{P}}{\bar z},\\
 c^{(\varsigma)}&\coloneqq\varsigma\frac{iK}{\sqrt{\mathcal Q}}
 +\frac{3{\sqrt{\mathcal Q}}}{\bar z}-\frac{3\partial_r \mathcal{Q}}{4{\sqrt{\mathcal Q}}}.
 \end{aligned}
 \label{eq:bg-chiral-bc}
\end{equation}

For any $2\times2$ matrix,
$\mathsf F^\#=(\operatorname{tr} \mathsf F) \mathsf I- \mathsf F$;
thus the tracelessness of $ \mathsf B^{(\varsigma)}$ gives
$\bigl(\mathsf F^{(\varsigma)}\bigr)^\#
=c^{(\varsigma)} \mathsf I+ \mathsf B^{(\varsigma)}$.
A dagger on a finite matrix, in contrast to the mode-operator
notation above, will mean conjugate transpose.

The matrix $ \mathsf A$ is Hermitian and depends only on the
background, whereas the mode parameters enter
$ \mathsf F^{(\varsigma)}$ through $K$ and $W$.
Their linear dependence on $\omega,m$ gives
\begin{equation}
  \mathsf F^{(-)}(\omega,m)= \mathsf F^{(+)}(-\omega,-m).
 \label{eq:bg-spin-reversal}
\end{equation}
This common matrix structure holds for both spin sectors and
for complex frequencies. It supplies the input to the
separation construction; the existence of the separating
transformation still has to be established.

For comparison with Kerr--Newman--de Sitter, take $\Lambda>0$ and $b=\ell={Q_m}=0$.
Then $\Omega=1$, $\mathcal M=M$, and the metric functions reduce
to their Kerr--Newman--de Sitter expressions.
The coordinate and mode normalizations of
Ref.~\cite{Yoshida:2026yly} are related to the present ones by
\begin{equation}
 t=\frac{t_{\mathrm{KNdS}}}{\Xi},\qquad
 \phi=\frac{\phi_{\mathrm{KNdS}}}{\Xi},\qquad
 \Xi\coloneqq 1+\frac{\Lambda a^2}{3},\qquad
 \omega=\Xi\omega_{\mathrm{KNdS}},\quad m=\Xi m_{\mathrm{KNdS}}.
 \label{eq:bg-KNdS-limit}
\end{equation}
For ${Q_e}={Q}>0$, this identifies the field normalizations and
matrices with those of that reference.
If ${Q_e}<0$, their signed-charge convention is recovered by a
constant sign change of the second chiral components.

\section{Application of the criterion to the PD system}
\label{sec:separated-equations}
We now apply the constructive criterion of
Sec.~\ref{sec:complete-criterion} to the PD chiral system.
Starting from its Hermitian sources, we determine a Gram decomposition
and construct a separated source frame. We then check the transport
conditions and reconstruct the separating matrix $\mathsf X$, from
which the radial and angular equations follow. Regularity, special
backgrounds, and comparison with earlier results are discussed after
the construction.

\subsection[Hermitian sources and Gram decomposition]{Hermitian sources and Gram decomposition}
\label{sec:pd-source-structure}
\label{app:separation-construction}

We begin with the coefficients of Sec.~\ref{sec:chiral-system}
on the background patch in Eq.~\eqref{eq:bg-local-domain}.
The first step in Theorem~\ref{thm:complete-criterion} is to compute
the source Gram matrix and find an additive decomposition with the
required ranks and signatures. For $\mathfrak m\coloneqq(\omega,m,\varsigma)$,
the coupling has the source representation

\begin{equation}
  \mathsf F^{(\varsigma)}
 =3 \mathsf H_0+3i \mathsf H_1
  +i\varsigma\bigl(l \mathsf T_l+\omega \mathsf T_\omega\bigr).
 \label{eq:pd-mode-matrix}
\end{equation}
Here the local north-chart mode combination is
\begin{equation}
 l\coloneqq am-(a+\ell)^2\omega,
 \label{eq:pd-mode-j}
\end{equation}
whose relation to global Fourier labels is given in
Sec.~\ref{sec:global-angular}.
The source matrices are
\begin{align}
  \mathsf H_0&\coloneqq \frac{ \mathsf F_0+ \mathsf F_0^\dagger}{6},&
  \mathsf H_1&\coloneqq \frac{ \mathsf F_0- \mathsf F_0^\dagger}{6i},\label{eq:pd-H-sources}\\
  \mathsf T_l&\coloneqq \frac{ \mathsf I}{{\sqrt{\mathcal Q}}}+\frac{\sigma_3}{\sqrt{P}},&
  \mathsf T_\omega&\coloneqq -\frac{r^2}{{\sqrt{\mathcal Q}}} \mathsf I
                       +\frac{p^2}{\sqrt{P}}\sigma_3,
 \label{eq:separation-sources}
\end{align}
where $ \mathsf F_0\coloneqq  \mathsf F^{(\varsigma)}|_{\omega=m=0}$.
These four Hermitian matrices are known from the background.
Thus the source family in Eq.~\eqref{eq:general-source-family} is
\begin{equation}
 ( \mathsf S_1, \mathsf S_2, \mathsf S_3, \mathsf S_4)
 \coloneqq ( \mathsf H_0, \mathsf H_1, \mathsf T_l, \mathsf T_\omega),
 \qquad
 d\coloneqq (3,3i,i\varsigma l,i\varsigma\omega).
 \label{eq:pd-source-family}
\end{equation}
Since $a>0$, the linear change from $(m,\omega)$ to $(l,\omega)$
is invertible.
For either fixed spin label, the two mode coefficients and the
Hermitian and anti-Hermitian parts of $ \mathsf F_0$ therefore
recover all four sources.
Thus the real mode family satisfies the source recoverability
assumption of Sec.~\ref{sec:coupling-construction}.
Taking Hermitian and anti-Hermitian parts commutes with the matrix
transformation in Eq.~\eqref{eq:direct-field-transformation},
so this argument does not require the separated scalar coefficients to be real.
Let $\Gamma$ be the Gram matrix of these four sources, ordered
as $(\mathsf H_0,\mathsf H_1,\mathsf T_l,\mathsf T_\omega)$.
To evaluate its entries, write $\mathcal M=\mu-in$ with $\mu,n$ real
and assume $|\mathcal M|>0$. The metric functions satisfy the following polynomial identities

\begin{align}
 \mathcal{Q}(r)&=k+{Q}^2-2\mu r+\epsilon r^2-2bnr^3
       -(b^2k+\frac{\Lambda}{3})r^4,
 \label{eq:separation-proof-radial-quartic}\\
 \mathcal P(\theta)&= \frac{1}{a^2\sin^2\theta} \left( k+2np-\epsilon p^2+2b\mu p^3
       -\bigl(b^2(k+{Q}^2)+\frac{\Lambda}{3}\bigr)p^4\right),
 \label{eq:separation-proof-quartics}
\end{align}
where

\begin{equation}
 k\coloneqq a^2-\ell^2,\qquad
 \epsilon\coloneqq
 1+4b\ell M-b^2k(k+{Q}^2)
 -\frac{\Lambda}{3}(a^2+3\ell^2).
 \label{eq:separation-proof-notation}
\end{equation}
\equationrefalias{eq:separation-proof-notation}{eq:pd-epsilon}

Direct evaluation with the bilinear form in
Eq.~\eqref{eq:separation-inner-product} gives the additive decomposition
\begin{equation}
 \Gamma(r,\theta)=\Gamma_r(r)+\Gamma_{\theta}(\theta).
 \label{eq:separation-gram-split}
\end{equation}
The freedom described in Eq.~\eqref{eq:general-constant-split-freedom}
is parametrized here by a real symmetric constant matrix,
\begin{equation}
 \mathsf K\coloneqq
 \begin{pmatrix}K_{00}&K_{01}\\K_{01}&K_{11}\end{pmatrix}.
 \label{eq:pd-gram-shift-matrix}
\end{equation}
Adding this matrix to the lower-right $2\times2$ radial block
and subtracting it from the angular block gives
\begin{align}
 \Gamma_r&\coloneqq \begin{pmatrix}
 h_{r}&0&x_r&y_r\\
 0&0&0&0\\
 x_r&0&\mathcal{Q}^{-1}+K_{00}&-r^2\mathcal{Q}^{-1}+K_{01}\\
 y_r&0&-r^2\mathcal{Q}^{-1}+K_{01}&r^4\mathcal{Q}^{-1}+K_{11}
 \end{pmatrix},\label{eq:pd-radial-gram}\\
 \Gamma_{\theta}&\coloneqq \begin{pmatrix}
 0&0&0&0\\
 0&h_{\theta}&x_{\theta}&y_{\theta}\\
 0&x_{\theta}&-P^{-1}-K_{00}&-p^2P^{-1}-K_{01}\\
 0&y_{\theta}&-p^2P^{-1}-K_{01}&-p^4P^{-1}-K_{11}
 \end{pmatrix}.
 \label{eq:separation-proof-gram-blocks}
\end{align}
This choice places $\mathsf H_0$ entirely in the radial part and
$\mathsf H_1$ entirely in the angular part.
The remaining entries are given by the source inner products:
\begin{align}
 h_{r}(r)\coloneqq\langle \mathsf H_0, \mathsf H_0\rangle
 &=\frac{(\partial_r \mathcal{Q})^2}{16\mathcal{Q}}+(\frac{\Lambda}{3}+b^2k)r^2+bnr-\frac{|\mathcal M|^2}{4{Q}^2},\label{eq:pd-hr}\\
 h_{\theta}(\theta)\coloneqq\langle \mathsf H_1, \mathsf H_1\rangle
 &=-\frac{(\sin\theta\,\partial_\theta\mathcal P+2\mathcal P\cos\theta)^2}{16\sin^2\theta\,\mathcal P}-\bigl(\frac{\Lambda}{3}+b^2(k+{Q}^2)\bigr)p^2+b\mu p-\frac{|\mathcal M|^2}{4{Q}^2},\label{eq:pd-htheta}\\
 \langle \mathsf H_0, \mathsf H_1\rangle&=0,
 \label{eq:separation-proof-source-norms}
\end{align}
and
\begin{align}
 x_r\coloneqq\langle \mathsf H_0, \mathsf T_l\rangle
 &=-\frac{\partial_r \mathcal{Q}}{4\mathcal{Q}},&
 y_r\coloneqq\langle \mathsf H_0, \mathsf T_\omega\rangle
 &=-r+\frac{r^2\partial_r \mathcal{Q}}{4\mathcal{Q}},\label{eq:pd-radial-cross-invariants}\\
 x_{\theta}\coloneqq\langle \mathsf H_1, \mathsf T_l\rangle
 &=-\frac{\sin\theta\,\partial_\theta\mathcal P+2\mathcal P\cos\theta}{4a\sin^2\theta\,\mathcal P},&
 y_{\theta}\coloneqq\langle \mathsf H_1, \mathsf T_\omega\rangle
 &=-p-\frac{p^2(\sin\theta\,\partial_\theta\mathcal P+2\mathcal P\cos\theta)}{4a\sin^2\theta\,\mathcal P}.
 \label{eq:separation-proof-cross-invariants}
\end{align}
The remaining inner products follow from the diagonal mode matrices:
\begin{equation}
 \begin{pmatrix}
 \langle \mathsf T_l, \mathsf T_l\rangle&
 \langle \mathsf T_l, \mathsf T_\omega\rangle\\
 \langle \mathsf T_\omega, \mathsf T_l\rangle&
 \langle \mathsf T_\omega, \mathsf T_\omega\rangle
 \end{pmatrix}
 =\frac1{\mathcal{Q}}\begin{pmatrix}1&-r^2\\-r^2&r^4\end{pmatrix}
  -\frac1{P}\begin{pmatrix}1&p^2\\p^2&p^4\end{pmatrix}.
 \label{eq:separation-proof-mode-gram}
\end{equation}
The required rank conditions reduce to
the vanishing of the principal $3\times3$ determinants obtained by
deleting the second row and column of $\Gamma_r$ and the first row
and column of $\Gamma_\theta$. Multiplying these determinants by
$16Q^2\mathcal Q$ and $16Q^2P$, respectively, gives polynomials.
Equating their coefficients to zero determines the three constants.
The independent conditions are

\begin{gather}
 K_{00}=a_2\det\mathsf K,\qquad K_{11}=-a_0\det\mathsf K,
 \label{eq:pd-gram-linear-conditions}\\
 \mu(K_{01}-a_1\det\mathsf K)=n(K_{01}-a_1\det\mathsf K)=0,
 \label{eq:pd-gram-offdiagonal-condition}\\
 1+(a_2a_0+a_1^2)\det\mathsf K=0,
 \label{eq:pd-gram-coefficient-conditions}
\end{gather}
where the background combinations appearing in the coefficients,
and the normalization used below, are

\begin{align}
 a_0&\coloneqq
 a^2-\ell^2+\frac{{Q}^2n^2}{|\mathcal M|^2},&
 a_2&\coloneqq\frac\Lambda3+b^2a_0,\label{eq:pd-a0-a2}\\
 a_1&\coloneqq
 \frac{|\mathcal M|^2-{Q}^2\epsilon}{2{Q}^2}
 +\frac{{Q}^2b\mu n}{|\mathcal M|^2},&
 \nu&\coloneqq
 2{Q}^2\sqrt{a_2a_0+a_1^2}.
 \label{eq:pd-construction-constants}
\end{align}
Assume $a_2a_0+a_1^2>0$ and take $\nu>0$.
Since $\mu^2+n^2>0$, the linear conditions give $K_{01}=a_1\det\mathsf K$.
The last condition then fixes $\det\mathsf K=-4Q^4/\nu^2$, and hence

\begin{equation}
 \mathsf K{}=\frac{4{Q}^4}{\nu^2}
 \begin{pmatrix}-a_2&-a_1\\-a_1&a_0\end{pmatrix},
 \label{eq:separation-gram-constant}
\end{equation}
\equationrefalias{eq:separation-gram-constant}{eq:separation-proof-constant-solution}
Substitution makes both determinant polynomials vanish
identically. We have thus obtained

\begin{equation}
 \operatorname{rank}\Gamma_r\leq2,\qquad
 \operatorname{rank}\Gamma_{\theta}\leq2.
 \label{eq:pd-gram-rank-bounds}
\end{equation}
We check that the four sources form a basis.
Their Gram determinant is
\begin{equation}
 \det\Gamma=-\frac{Z_{\rm src}^2}{16{Q}^4\mathcal{Q}P},\qquad
 Z_{\rm src}\coloneqq |\mathcal M|^2(r^2+p^2)-2{Q}^2(1+brp)(\mu r+np)+4{Q}^4brp.
 \label{eq:separation-proof-source-determinant}
\end{equation}
We work on an open rectangle where $Z_{\rm src}\ne0$.
The Gram matrix is then nonsingular, so the four sources are
linearly independent.
The rank bounds established above now imply
\begin{equation}
 4=\operatorname{rank}\Gamma
 \leq\operatorname{rank}\Gamma_r+\operatorname{rank}\Gamma_\theta
 \leq4.
 \label{eq:pd-gram-rank-equality}
\end{equation}
Hence both $\Gamma_r$ and $\Gamma_\theta$ have rank two.

It remains to check the signatures required by
Theorem~\ref{thm:coupling-construction}. The $2\times2$ principal
submatrix of $\Gamma_r$ corresponding to $\mathsf T_l$ and
$\mathsf T_\omega$ has determinant
\begin{equation}
 \det\begin{pmatrix}
 \mathcal Q^{-1}+K_{00}&-r^2\mathcal Q^{-1}+K_{01}\\
 -r^2\mathcal Q^{-1}+K_{01}&r^4\mathcal Q^{-1}+K_{11}
 \end{pmatrix}
 =-\frac{4{Q}^2(|\mathcal M|^2 r-{Q}^2\mu-b{Q}^2nr^2)^2}{|\mathcal M|^2\nu^2\mathcal{Q}}.\label{eq:pd-radial-mode-minor}
\end{equation}
This determinant is negative away from isolated radial zeros,
so this submatrix has one positive and one negative eigenvalue.
Hence $\Gamma_r$ has both positive and negative directions.
Since its rank is two, the radial plane has signature $(+,-)$.
The full source space has signature $(+,-,-,-)$,
so the orthogonal angular plane is negative definite.
These signatures persist across the isolated zeros because both
planes remain nondegenerate wherever $\Gamma$ is nonsingular.

This verifies the hypotheses of Theorem~\ref{thm:coupling-construction}.
We next implement its source-frame construction explicitly.
The additional assumptions under which this Gram decomposition is
unique are stated and analyzed in Sec.~\ref{app:pd-rank-comparison};
they are not needed to construct the decomposition above.

\subsection[Construction of the source frame and verification of the separation conditions]{Construction of the source frame and verification of the separation conditions}
\label{sec:pd-matrix-construction}

The Gram decomposition determines two orthogonal planes in
the space of Hermitian matrices. We construct $\mathsf W$ and $\mathsf E$
directly using the criterion in
Eqs.~\eqref{eq:general-direct-frame-normalization} and
\eqref{eq:general-direct-frame-separation}, using $r$ and
$\theta$ as the separate radial and angular variables.
We first choose normalized projections in these planes, then verify
positivity and the required one-variable dependence.
The projections can be computed directly from the sources:

\begin{equation}
 \Pi_r(\mathsf S_a)\coloneqq
\sum_{b=1}^4\mathsf S_b(\Gamma^{-1}\Gamma_r)_{ba},\qquad
\Pi_\theta(\mathsf S_a)\coloneqq
\mathsf S_a-\Pi_r(\mathsf S_a).
 \label{eq:pd-source-projections}
\end{equation}
Indeed, in the factorization of
Eq.~\eqref{eq:general-gram-factors}, $\Gamma^{-1}\Gamma_r$ is similar
to $\operatorname{diag}(1,1,0,0)$. It is an orthogonal projection
for the source Gram form. Thus the inner products of projected
sources are the entries of $\Gamma_r$ and $\Gamma_\theta$.
For the explicit source-frame construction, we introduce the functions

\begin{equation}
 F_r(r)\coloneqq
 \frac{|\mathcal M|^2r-{Q}^2\mu-b{Q}^2nr^2}{|\mathcal M|},
 \qquad
 F_{\theta}(\theta)\coloneqq
 \frac{|\mathcal M|^2p-{Q}^2n-b{Q}^2\mu p^2}{|\mathcal M|}.
 \label{eq:pd-Fr-Fp}
\end{equation}

To construct $\mathsf W$ and $\mathsf E$, we choose a linear
combination of the source matrices $\mathsf T_l$ and $\mathsf T_\omega$.
Its radial and angular projections will be normalized to define
these two frame matrices. We consider
\begin{equation}
 \mathsf T_\kappa\coloneqq
\kappa\mathsf T_l-a_2\mathsf T_\omega,
 \label{eq:pd-normalizing-source}
\end{equation}
where $\kappa$ is a real constant to be fixed below.
The determinant of the radial projection of $\mathsf T_\kappa$ is
\begin{equation}
 \det\Pi_r(\mathsf T_\kappa)
 =\frac{(\kappa+a_2r^2)^2}{\mathcal Q}
 +\begin{pmatrix}\kappa&-a_2\end{pmatrix}\mathsf K
 \binom{\kappa}{-a_2}.
 \label{eq:pd-radial-projected-norm-general}
\end{equation}
To simplify its normalization, we choose $\kappa$ so that
the second term vanishes. This term is

\begin{equation}
 \begin{pmatrix}\kappa&-a_2\end{pmatrix}\mathsf K
\binom{\kappa}{-a_2}
=-\frac{4Q^4a_2}{\nu^2}
 (\kappa^2-2a_1\kappa-a_2a_0).
 \label{eq:pd-normalizing-constant}
\end{equation}
We therefore set

\begin{equation}
 \kappa_\star\coloneqq a_1+\sqrt{a_1^2+a_2a_0},\qquad
 \kappa_\star^2-2a_1\kappa_\star-a_2a_0=0,\qquad
 \frac{\nu}{{Q}^2}=2(\kappa_\star-a_1).
 \label{eq:pd-kappa-star}
\end{equation}
The same choice applies at $a_2=0$ by continuity.
For this representative, restrict the background patch further to

\begin{equation}
 \kappa_\star>0,\qquad
 \kappa_\star+a_2r^2>0,\qquad
 \kappa_\star-a_2p^2>0,\qquad F_r>{Q}{\sqrt{\mathcal Q}}.
 \label{eq:pd-construction-patch}
\end{equation}
The positive square roots will be used throughout.
The source projections then have the simple norms

\begin{equation}
 \det\Pi_r(\mathsf T_{\kappa_\star})
=\frac{(\kappa_\star+a_2r^2)^2}{\mathcal Q},\qquad
\det\Pi_\theta(\mathsf T_{\kappa_\star})
=-\frac{(\kappa_\star-a_2p^2)^2}{P}.
 \label{eq:pd-projected-norms}
\end{equation}
Normalizing the radial and angular projections of $\mathsf T_{\kappa_\star}$,
we set

\begin{equation}
 \mathsf W\coloneqq
\frac{\sqrt{\mathcal Q}}{\kappa_\star+a_2r^2}
 \Pi_r(\mathsf T_{\kappa_\star}),\qquad
\mathsf E\coloneqq
-\frac{\sqrt{P}}{\kappa_\star-a_2p^2}
 \Pi_\theta(\mathsf T_{\kappa_\star}).
 \label{eq:pd-normalized-WE}
\end{equation}
These matrices are Hermitian, with $\det\mathsf W=1$
and $\det\mathsf E=-1$. Evaluating the trace from the source projection gives

\begin{equation}
 \frac12\operatorname{tr}\mathsf W
=1+\frac{2Q^2\mathcal Q(\kappa_\star-a_2p^2)}
 {\nu\Sigma(\kappa_\star+a_2r^2)}>0,
 \label{eq:pd-W-positive}
\end{equation}
so $\mathsf W$ is positive definite.
This verifies Eq.~\eqref{eq:general-direct-frame-normalization}.
To check Eq.~\eqref{eq:general-direct-frame-separation}, note that
the inner products of the projected source $\mathsf T_{\kappa_\star}$
with $\mathsf S_a$ are constant linear combinations of entries of
the corresponding Gram block. The normalization of $\mathsf W$
depends only on $r$, and that of $\mathsf E$ only on $\theta$.
Hence $\langle\mathsf W,\mathsf S_a\rangle$ depends only on $r$,
and $\langle\mathsf E,\mathsf S_a\rangle$ only on $\theta$.
The direct construction criterion therefore guarantees corresponding
factors $\mathsf R(r)$ and $\mathsf Q(\theta)$.

Evaluating the projections in Eq.~\eqref{eq:pd-normalized-WE}, we obtain
\begin{align}
 W_{11,22}&=1+\frac{2Q^2\sqrt{\mathcal Q}}
 {\nu\Sigma}(s^2\sqrt{\mathcal Q}\mathbin{\pm}\sqrt{P}),
 \label{eq:pd-W-diagonal}\\
 W_{12}&=\frac{2Q\sqrt{\mathcal Q}\,\mathcal M}
 {\nu|\mathcal M|z^2}(F_\theta-is^2F_r),\qquad
 W_{21}=\overline{W_{12}},
 \label{eq:pd-W-offdiagonal}\\
 E_{11,22}&=\mathbin{\mp}1+\frac{2Q^2\sqrt{P}}
 {\nu\Sigma}(\sqrt{\mathcal Q}\mathbin{\pm}s^{-2}\sqrt{P}),
 \label{eq:pd-E-diagonal}\\
 E_{12}&=\frac{2Q\sqrt{P}\mathcal M}{\nu|\mathcal M|z^2}
 (s^{-2}F_\theta-iF_r),\qquad E_{21}=\overline{E_{12}},
 \label{eq:pd-WE-components}
\end{align}
where
\begin{equation}
 s(r,\theta)\coloneqq
 \sqrt{\frac{\kappa_\star-a_2p^2}{\kappa_\star+a_2r^2}}.
 \label{eq:pd-reconstruction-ratio}
\end{equation}
The upper and lower signs refer to the $11$ and $22$ entries,
respectively.

We now complete the frame. To fix the radial orientation, take
$\mathsf U\coloneqq\Pi_r(\mathsf T_l)$ and
$\mathsf V\coloneqq\Pi_r(\mathsf T_\omega)$ in
Eq.~\eqref{eq:separation-J-reconstruction}. On the chosen branch this gives

\begin{equation}
 \mathsf J\coloneqq
-\frac{\mathsf U\mathsf V^\#
 -\langle\mathsf U,\mathsf V\rangle\mathsf I}{d_{UV}},\qquad
d_{UV}=\frac{2QF_r}{\nu\sqrt{\mathcal Q}}>0.
 \label{eq:pd-J-from-projections}
\end{equation}
To evaluate this expression, we use the polynomial identities
\begin{align}
 F_r^2-{Q}^2{\mathcal Q}
 &={Q}^2\left(a_2r^4
       +2a_1r^2-a_0\right),\label{eq:pd-radial-block-identity}\\
 F_{\theta}^2+{Q}^2P
 &={Q}^2\left(-a_2p^4
       +2a_1p^2+a_0\right).
 \label{eq:pd-block-identities}
\end{align}
Substituting the projected sources into
Eq.~\eqref{eq:pd-J-from-projections} and using these identities gives
\begin{align}
 \mathsf J_{11}=-\mathsf J_{22}
 &=\frac{2{Q}(F_r\sqrt{P}-iF_{\theta}{\sqrt{\mathcal Q}})}{\nu\Sigma},\label{eq:pd-J-diagonal}\\
 \mathsf J_{12}
 &=\frac{i\mathcal M \bar z}{|\mathcal M|z\nu\Sigma}
 \left({Q}^2({\sqrt{\mathcal Q}}+\sqrt{P})^2-(F_r+iF_{\theta})^2
       +{Q}^2a_2\Sigma^2\right),\label{eq:pd-J-upper}\\
 \mathsf J_{21}
 &=-\frac{i\overline{\mathcal M}z}{|\mathcal M|{\bar z}\nu\Sigma}
 \left({Q}^2({\sqrt{\mathcal Q}}-\sqrt{P})^2-(F_r-iF_{\theta})^2
       +{Q}^2a_2\Sigma^2\right).
 \label{eq:pd-J-explicit}
\end{align}

The matrices constructed above give the separated source frame
\begin{equation}
 (\mathsf B_0,\mathsf B_1,\mathsf B_2,\mathsf B_3)
 \coloneqq(\mathsf W,\mathsf J\mathsf W,i\mathsf J\mathsf E,\mathsf E).
 \label{eq:pd-source-frame}
\end{equation}
By Theorem~\ref{thm:coupling-construction}, this frame determines
a local transformation satisfying the coupling separation condition.
We now verify the remaining transport conditions.
Using Eq.~\eqref{eq:source-frame-transport}, we obtain

\begin{equation}
 \begin{aligned}
\mathsf H_F
&=\frac14\sum_{i=0}^3\eta_i
 (\mathscr D\mathsf B_i)\mathsf B_i^\#\\
&=Q\Omega\begin{pmatrix}0&-i\bar z/z^2\\iz/\bar z^2&0\end{pmatrix}
 +(\xi(r)+\eta(\theta))\mathsf J,
\qquad (\eta_0,\eta_1,\eta_2,\eta_3)=(1,-1,-1,-1),
\end{aligned}
 \label{eq:pd-source-frame-transport}
\end{equation}
where

\begin{equation}
 \begin{aligned}
 \xi&=-\frac{b{Q}nr}{|\mathcal M|}
   -\frac{a_2rF_r}{{Q}(\kappa_\star+a_2r^2)},\\
 \eta&=\frac{b{Q}\mu p}{|\mathcal M|}
   -\frac{a_2pF_{\theta}}{{Q}(\kappa_\star-a_2p^2)}.
 \end{aligned}
 \label{eq:pd-Xa-transport-coefficients}
\end{equation}
The first matrix is precisely the Hermitian coefficient
$\mathsf A$ in Eq.~\eqref{eq:bg-chiral-A}; $\xi$ and $\eta$ are real
and depend only on their indicated coordinates.
Hence $\mathsf H_F-\mathsf A=(\xi+\eta)\mathsf J$, which verifies
all three transport tests in Eq.~\eqref{eq:complete-transport-condition}.
The evaluation follows by direct differentiation of the
frame components in
Eqs.~\eqref{eq:pd-W-diagonal}--\eqref{eq:pd-WE-components}
and \eqref{eq:pd-J-diagonal}--\eqref{eq:pd-J-explicit},
using the background quartics.
Theorem~\ref{thm:complete-criterion} therefore establishes local separation.

On the source-basis patch, Theorem~\ref{thm:hermitian-transport}
also establishes uniqueness of the compatible Hermitian coefficient
for the specified differential operator, Gram decomposition, and
oriented frame. Indeed,
$\operatorname{Im}\mathsf J_{11}=-2QF_\theta\sqrt{\mathcal Q}/(\nu\Sigma)$
is nonzero on a dense open subset. The compatible coefficient is
therefore the original $\mathsf A$, with the equality extending
by smoothness.

\subsection[Construction of the separating matrix]{Construction of the separating matrix}
\label{sec:pd-X-construction}
\label{sec:transformation-properties}

Having established local separation, we now construct the
separating matrix $\mathsf X$ explicitly from the source frame.
As in Sec.~\ref{sec:general-separation-theorem}, the axis
$\mathsf T=\tau_3$ can be chosen by a constant unitary change of basis.
The radial coupling in
Refs.~\cite{Hintz:2026jiw,Yoshida:2026yly,Zhao:2026gen,Gray:2026rzu}
is a linear combination of $\mathsf I$ and $\sigma_2$,
corresponding to the choice $\mathsf T=\sigma_2$ in our notation.
Here we incorporate the phase of
$\mathcal M$ into this axis to obtain the compact expression
for $\mathsf X$ below. We choose

\begin{equation}
 \mathsf T\coloneqq \frac{\mu\sigma_2-n\sigma_1}{|\mathcal M|}
 =\frac1{|\mathcal M|}
 \begin{pmatrix}0&-i\mathcal M\\i\overline{\mathcal M}&0\end{pmatrix},
 \qquad \mathsf T^2=\mathsf I,\quad \mathsf T^\dagger=\mathsf T,\quad\{\mathsf T,\sigma_3\}=0.
 \label{eq:pd-axis-T}
\end{equation}
This choice reduces to $\mathsf T=\sigma_2$ when $n=0$ and $\mu>0$.
Fix the eigenvector phases by
\begin{equation}
 e_+\coloneqq \frac1{\sqrt2}\binom{-i\mathcal M/|\mathcal M|}{1},\qquad
 e_-\coloneqq \frac1{\sqrt2}\binom{-\mathcal M/|\mathcal M|}{i}.
 \label{eq:pd-T-eigenvectors}
\end{equation}
They are orthonormal and obey $\mathsf Te_\pm=\pm e_\pm$ and
$e_-=(i\mathsf T\sigma_3)e_+$. Thus the adapted Pauli triple is
$(\tau_1,\tau_2,\tau_3)=(i\mathsf T\sigma_3,\sigma_3,\mathsf T)$.

The quartics in
Eqs.~\eqref{eq:separation-proof-radial-quartic}--\eqref{eq:separation-proof-quartics}
and the equation for $\kappa_\star$ give the factorizations
\begin{align}
 F_r^2-{Q}^2{\mathcal Q}
 &=\frac{{Q}^2}{\kappa_\star}
   (\kappa_\star r^2-a_0)(\kappa_\star+a_2r^2),
 \label{eq:pd-radial-factorization}\\
 F_{\theta}^2+{Q}^2P
 &=\frac{{Q}^2}{\kappa_\star}
   (a_0+\kappa_\star p^2)(\kappa_\star-a_2p^2).
 \label{eq:pd-Phi-factorization}
\end{align}
Since $\mathsf W-\mathsf E$ has rank one, its nonzero columns
are proportional. Rescaling $\beta$ changes the determinant-normalized
matrix $\mathsf X$ only by a sign. We therefore choose
\begin{equation}
 \beta\coloneqq
 \begin{pmatrix}
 \displaystyle\frac{\mathcal M}{|\mathcal M|z}(sF_r+is^{-1}F_\theta)\\[5pt]
 \displaystyle\frac{iQ}{\bar z}(s\sqrt{\mathcal Q}-s^{-1}\sqrt{P})
 \end{pmatrix}.
 \label{eq:pd-explicit-beta}
\end{equation}
Using these identities, the component expressions give
\begin{equation}
 \mathsf W-\mathsf E=\frac2\nu\beta\beta^\dagger.
 \label{eq:pd-reconstruction-column}
\end{equation}
To evaluate the reconstruction formula, we set
\begin{equation}
 \mathsf D_z\coloneqq\operatorname{diag}(z,\bar z).
 \label{eq:pd-Dz-definition}
\end{equation}
Substituting this vector and the chosen eigenvectors into
Eq.~\eqref{eq:normalization-proof-column-matrix} gives
\begin{align}
 \mathsf B
 &=(\mathsf I+\mathsf J)\beta e_+^\dagger
 +i(\mathsf I-\mathsf J)\beta e_-^\dagger
 \label{eq:pd-B-substitution}\\
 &=i\sqrt2\,\mathsf D_z^{-1}
 \bigl(s(Q\sqrt{\mathcal Q}\,\mathsf I+F_r\mathsf T)
 +s^{-1}(Q\sqrt{P}\mathsf I-iF_\theta\mathsf T)\sigma_3\bigr),
 \label{eq:pd-B-evaluated}\\
 \det\mathsf B&=\frac2\Sigma\bigl(s^2(F_r^2-Q^2\mathcal Q)
 +s^{-2}(F_\theta^2+Q^2P)\bigr)=2\nu.
 \label{eq:pd-reconstruction-evaluated}
\end{align}
Normalizing by $\sqrt{2\nu}$, we obtain

\begin{equation}
 \begin{aligned}
 \mathsf X{}= {}&\frac{i}{\sqrt\nu}
 \begin{pmatrix}z^{-1}&0\\0&\bar z^{-1}\end{pmatrix}
 \biggl(
 \sqrt{\frac{\kappa_\star-a_2p^2}{\kappa_\star+a_2r^2}}
       ({Q}{\sqrt{\mathcal Q}}\mathsf I+F_r\mathsf T)\\[-2pt]
 &\hspace{27mm}+
 \sqrt{\frac{\kappa_\star+a_2r^2}{\kappa_\star-a_2p^2}}
       ({Q}\sqrt{P}\mathsf I-iF_{\theta}\mathsf T)\sigma_3
 \biggr).
 \end{aligned}
 \label{eq:pd-Xa}
\end{equation}
\equationrefalias{eq:pd-Xa}{eq:pd-X-explicit}
Equivalently, its components are
\begin{equation}
 \mathsf X=\frac{i}{\sqrt\nu}
 \begin{pmatrix}
 \displaystyle\frac{{Q}}{z}\left(s{\sqrt{\mathcal Q}}+\frac{\sqrt{P}}{s}\right)
 &\displaystyle-\frac{i\mathcal M}{|\mathcal M|z}
     \left(sF_r+i\frac{F_{\theta}}{s}\right)\\[8pt]
 \displaystyle\frac{i\overline{\mathcal M}}{|\mathcal M|\bar z}
     \left(sF_r-i\frac{F_{\theta}}{s}\right)
 &\displaystyle\frac{{Q}}{\bar z}\left(s{\sqrt{\mathcal Q}}-\frac{\sqrt{P}}{s}\right)
 \end{pmatrix}.
 \label{eq:pd-X-components}
\end{equation}
Here the sign of the local square root is chosen to give the
displayed representative.

\paragraph{Remarks on regularity and boundary limits.}
We record three properties of the explicit separating matrix
$\mathsf X$: its extension through zeros of $Z_{\rm src}$,
its horizon limits, and its behavior at the angular endpoints.
\begin{enumerate}
\item \textit{Local regularity.}
The source-frame construction uses $\Gamma^{-1}$ and hence
assumes nonsingular sources. Direct substitution of
Eq.~\eqref{eq:pd-Xa} verifies $\det\mathsf X=1$ and the
transport and coupling identities, including at zeros of
$Z_{\rm src}$ whenever Eq.~\eqref{eq:pd-construction-patch}
holds. Thus the separation extends through these zeros.
This extension does not assert uniqueness of the Gram
decomposition there.

\item \textit{Horizon limits.}\label{app:pd-horizon-unitarity}
At a finite root $r_h$ of $\mathcal Q$ approached from the construction
patch, assume $\kappa_\star+a_2r_h^2>0$ and $F_r(r_h)>0$.
The denominators of Eq.~\eqref{eq:pd-Xa} are then nonzero at an interior
angle, so setting $\mathcal Q=0$ gives a finite matrix. In its numerator,
$\mathsf B\coloneqq ({Q}\sqrt{P}\mathsf I-iF_{\theta}\mathsf T)\sigma_3$ satisfies
$\mathsf B^\dagger=\mathsf B$, $\{\mathsf T,\mathsf B\}=0$, and
$\mathsf B^2=(F_\theta^2+Q^2P)\mathsf I$. Consequently, direct multiplication gives
\begin{equation}
 \mathsf X(r_h,\theta)
 \mathsf X(r_h,\theta)^\dagger=\mathsf I,\qquad
 \det\mathsf X(r_h,\theta)=1.
 \label{eq:pd-X-horizon-limit}
\end{equation}
Physical horizon conditions require the separate normalization of the
original radiative fields.

\item \textit{Angular endpoints.}
Fix $r$ in the construction patch. At the angular endpoints assume
\begin{equation}
 \mathcal P>0,\qquad a_0+\kappa_\star p^2>0
 \quad\hbox{on }[0,\pi].
 \label{eq:pd-endpoint-domain}
\end{equation}
Equations~\eqref{eq:pd-radial-factorization}--\eqref{eq:pd-Phi-factorization} then make
$\kappa_\star-a_2p^2$ nonnegative at each pole. If it is positive,
Eq.~\eqref{eq:pd-Xa} is manifestly smooth there. If it vanishes at a
pole, the same identity gives $F_{\theta}=0$ there. With
$x_N\coloneqq \theta$, $x_S\coloneqq \pi-\theta$, the background functions obey
\[
 F_{\theta}=O(x_A^2),\qquad \sqrt{P}=c_Ax_A+O(x_A^3),\qquad c_A>0,
 \qquad\kappa_\star-a_2p^2=c'_Ax_A^2+O(x_A^4),\quad c'_A>0.
\]
The first term in the brackets of Eq.~\eqref{eq:pd-Xa} is then
$O(x_A)$, and the second has a finite smooth one-sided limit:
its numerator ${Q}\sqrt{P}\mathsf I-iF_{\theta}\mathsf T$ is $O(x_A)$.
Thus $\mathsf X$ extends to the poles. Its determinant
remains one by continuity, so its inverse also extends and is bounded.
These statements concern chiral variables and do not select a physical
condition at a string or strut.
\end{enumerate}

\subsection{Separated coupling and component equations}
\label{sec:pd-Xa-equations}
We now read the coupling coefficients from the separated
source frame and apply Theorem~\ref{thm:chiral-separation}.
Extending the bilinear form complex linearly, the coefficients follow from

\begin{equation}
 \begin{aligned}
c_0&=\langle\mathsf B_0,\mathsf F^{(\varsigma)}\rangle,&
c_3&=-\langle\mathsf B_1,\mathsf F^{(\varsigma)}\rangle,\\
i\gamma_1&=-\langle\mathsf B_2,\mathsf F^{(\varsigma)}\rangle,&
i\gamma_2&=-\langle\mathsf B_3,\mathsf F^{(\varsigma)}\rangle.
\end{aligned}
 \label{eq:pd-coupling-from-frame}
\end{equation}
For each $\varsigma=\pm1$, the transformed coupling has the form
\begin{equation}
 \mathsf C=\mathsf X^{-1}\mathsf F^{(\varsigma)}(\mathsf X^\dagger)^{-1}
 =c_0\mathsf I+c_3\mathsf T
       -(\gamma_1\mathsf T-i\gamma_2\mathsf I)\sigma_3,
 \label{eq:pd-Ca}
\end{equation}
with radial coefficients
\begin{align}
 c_0={}&-\frac32\partial_r {\sqrt{\mathcal Q}}
 +\frac{3a_2r{\sqrt{\mathcal Q}}}{\kappa_\star+a_2r^2}
 +i\varsigma\left(
 \frac{l-\omega r^2}{{\sqrt{\mathcal Q}}}
 -\frac{2{Q}^2{\sqrt{\mathcal Q}}}{\nu}\frac{\kappa_\star\omega+a_2l}{\kappa_\star+a_2r^2}
 \right),\label{eq:pd-ca0}\\
 c_3={}&3\left(\frac{|\mathcal M|}{2{Q}}+\xi\right)
 +\frac{2i\varsigma {Q}F_r}{\nu}
     \frac{\kappa_\star\omega+a_2l}{\kappa_\star+a_2r^2},
 \label{eq:pd-ca3}
\end{align}
and angular coefficients
\begin{align}
 \gamma_1={}&3\left(\eta-\frac{|\mathcal M|}{2{Q}}\right)
 -\frac{2\varsigma {Q}F_{\theta}}{\nu}
     \frac{\kappa_\star\omega+a_2l}{\kappa_\star-a_2p^2},
 \label{eq:pd-ga1}\\
 \gamma_2={}&-\frac32\bigl(\partial_\theta\sqrt{\mathcal P}+\sqrt{\mathcal P}\cot\theta\bigr)
 +\frac{3a_2p\sqrt{P}}{\kappa_\star-a_2p^2}
 -\varsigma\left(
 \frac{l+\omega p^2}{\sqrt{P}}
 -\frac{2{Q}^2\sqrt{P}}{\nu}\frac{\kappa_\star\omega+a_2l}{\kappa_\star-a_2p^2}
 \right).\label{eq:pd-ga2}
\end{align}
The spin-sector label on these coefficients is suppressed. They are
algebraic in the mode labels, so the identities also hold for complex
$\omega,m$.

In the eigenbasis fixed in Eq.~\eqref{eq:pd-T-eigenvectors},

\begin{equation}
 \mathsf U_{\mathsf T}^\dagger \mathsf C\mathsf U_{\mathsf T}=
 \begin{pmatrix}
 c_0+c_3&\gamma_2+i\gamma_1\\
 -\gamma_2+i\gamma_1&c_0-c_3
 \end{pmatrix}.
 \label{eq:pd-Ca-components}
\end{equation}
The general separation theorem therefore gives the angular equations
\begin{equation}
 \begin{aligned}
 -i\sqrt{\mathcal P}\partial_\theta S_1+\eta S_1
       -(\gamma_2+i\gamma_1)S_2
       &=\zeta S_1,\\
 i\sqrt{\mathcal P}\partial_\theta S_2+\eta S_2
       +(-\gamma_2+i\gamma_1)S_1
       &=\zeta S_2,
 \end{aligned}
 \label{eq:pd-Xa-angular}
\end{equation}
and the radial equations
\begin{equation}
 \begin{aligned}
 {\sqrt{\mathcal Q}}\partial_r R_1&=-(\zeta+\xi)R_1
                   +(c_0+c_3)R_2,\\
 {\sqrt{\mathcal Q}}\partial_r R_2&=(c_0-c_3)R_1
                   +(\zeta+\xi)R_2.
 \end{aligned}
 \label{eq:pd-Xa-radial}
\end{equation}
The reconstruction of the chiral variables is explicitly
\begin{equation}
 \begin{aligned}
 y_+^{(\varsigma)}&=\mathsf X
       (R_1S_1e_++R_2S_2e_-),\\
 y_-^{(\varsigma)}&=(\mathsf X^\dagger)^{-1}
       (R_2S_1e_++R_1S_2e_-).
 \end{aligned}
 \label{eq:pd-Xa-fields}
\end{equation}
The angular operator is
\begin{equation}
 \mathcal L=-i\sqrt{\mathcal P}\sigma_3\partial_\theta
             -\gamma_2\sigma_1
             +\gamma_1\sigma_2+\eta\mathsf I.
 \label{eq:pd-La}
\end{equation}
The $\sigma_i$ here act on the component column $(S_1,S_2)^T$;
the axis $\mathsf T$ acts on the original internal doublet.

\subsection{Special backgrounds and comparison with earlier results}
\label{sec:pd-special-backgrounds}
Having constructed the PD separating matrix and examined its
regularity, we now specialize the background parameters and compare
the resulting transformations with earlier results.

\paragraph{The algebraic branch $a_2=0$.}
The condition $a_2=0$ is equivalent to $\Lambda=-3b^2a_0$.
On the construction branch with $\kappa_\star>0$, the square-root
ratios disappear:
\begin{equation}
 \mathsf X=\frac{i}{\sqrt\nu}
 \begin{pmatrix}
 {Q}({\sqrt{\mathcal Q}}+\sqrt{P})/z&-i\mathcal M(F_r+iF_{\theta})/(|\mathcal M|z)\\
 i\overline{\mathcal M}(F_r-iF_{\theta})/(|\mathcal M|\bar z)&{Q}({\sqrt{\mathcal Q}}-\sqrt{P})/\bar z
 \end{pmatrix}.
 \label{eq:pd-Xa-a2zero}
\end{equation}
Moreover $\xi=-b{Q}nr/|\mathcal M|$ and
$\eta=b{Q}\mu p/|\mathcal M|$, so the transport identity becomes
\[
 \mathscr D\mathsf X=\mathsf A\mathsf X
 +\frac{b{Q}(\mu p-nr)}{|\mathcal M|}\mathsf X\mathsf T.
\]
Thus $a_2=0$ simplifies the matrix but does not in general make
its transport residual vanish.

\paragraph{Kerr--Newman--NUT and Kerr--Newman limits.}
Setting $b=\Lambda=0$ gives the Kerr--Newman--NUT background,
with electric and magnetic charges allowed. Then $\mu=M$, $n=\ell$,
and $a_2=0$. On the branch $|\mathcal M|^2>{Q}^2$,
$\nu=|\mathcal M|^2-{Q}^2$, and the matrix and its transport reduce to
\begin{equation}
 \mathsf X=\frac{i}{\sqrt{|\mathcal M|^2-{Q}^2}}
 \begin{pmatrix}
 {Q}({\sqrt{\mathcal Q}}+\sqrt{P})/z&-i(\mathcal M-{Q}^2/z)\\
 i(\overline{\mathcal M}-{Q}^2/\bar z)&{Q}({\sqrt{\mathcal Q}}-\sqrt{P})/\bar z
 \end{pmatrix},\qquad \mathscr D\mathsf X=\mathsf A\mathsf X.
 \label{eq:pd-Xa-flat}
\end{equation}
The vanishing transport residual here concerns this explicit
representative of the same separation planes.

To compare with Hintz's integrating factor~\cite{Hintz:2026jiw},
we further set $\ell={Q_m}=0$ and take
$M>0$, $a>0$, ${Q_e}={Q}>0$, and
$M^2>a^2+{Q}^2$. On the exterior $r>r_+=M+\sqrt{M^2-a^2-{Q}^2}$,
we have $\nu=M^2-{Q}^2$, $\mathcal Q=r^2-2Mr+a^2+{Q}^2$,
$\mathcal P=1$, and $z=r+ia\cos\theta$.
With $\bar\rho=z$ and $\bar\rho^*=\bar z$ in that reference's notation,
its matrix integrating factor in
Eq.~(5) is
\begin{equation}
 \mathsf X_{\mathrm H}=
 \begin{pmatrix}
 {Q}({\sqrt{\mathcal Q}}+\sqrt{P})/z&-i(M-{Q}^2/z)\\
 i(M-{Q}^2/\bar z)&{Q}({\sqrt{\mathcal Q}}-\sqrt{P})/\bar z
 \end{pmatrix},\qquad
 \det\mathsf X_{\mathrm H}=-\nu.
 \label{eq:pd-KN-Hintz-matrix}
\end{equation}
Equation~\eqref{eq:pd-Xa-flat} therefore gives the direct relation
\begin{equation}
 \mathsf X=\frac{i}{\sqrt\nu}\mathsf X_{\mathrm H}.
 \label{eq:pd-KN-Hintz-X}
\end{equation}
The factor $i/\sqrt\nu$ normalizes Hintz's matrix to unit determinant.
Here $\xi=\eta=0$, so the transport identity reduces to
$\mathscr D\mathsf X=\mathsf A\mathsf X$, reproducing the
integrating-factor identity in Eq.~(6) of that reference.

\paragraph{Kerr--Newman--de Sitter limit and comparison.}
For $\Lambda>0$, $b=\ell={Q_m}=0$, $M>0$, and ${Q_e}={Q}>0$, one has
$\mathcal M=M$, $\mathsf T=\sigma_2$, $F_r=Mr-{Q}^2$, $F_{\theta}=Mp$,
and $a_2=\Lambda/3$. The constants reduce to
\begin{equation}
 \begin{aligned}
 \nu&=\sqrt{\Biggl(M^2-{Q}^2\left(1-\frac{\Lambda a^2}{3}\right)\Biggr)^2
              +\frac{4\Lambda a^2{Q}^4}{3}},\\
 \kappa_\star&=\frac{M^2-{Q}^2(1-\Lambda a^2/3)+\nu}{2{Q}^2}.
 \end{aligned}
 \label{eq:pd-KNdS-constants}
\end{equation}
In this limit, Eq.~\eqref{eq:pd-X-components} becomes
\begin{equation}
 \mathsf X=\frac1{\sqrt\nu}
 \begin{pmatrix}
 \displaystyle\frac{iQ}{z}
   \left(s\sqrt{\mathcal Q}+\frac{\sqrt{P}}{s}\right)
 &
 \displaystyle\frac{s(Mr-Q^2)+iMp/s}{z}
 \\[8pt]
 \displaystyle-\frac{s(Mr-Q^2)-iMp/s}{\bar z}
 &
 \displaystyle\frac{iQ}{\bar z}
   \left(s\sqrt{\mathcal Q}-\frac{\sqrt{P}}{s}\right)
 \end{pmatrix},
 \label{eq:pd-KNdS-X-explicit}
\end{equation}
where
\begin{equation}
 s=\sqrt{\frac{\kappa_\star-\Lambda p^2/3}
                    {\kappa_\star+\Lambda r^2/3}},
 \qquad p=a\cos\theta,\qquad z=r+ip.
 \label{eq:pd-KNdS-X-functions}
\end{equation}
The transport functions and separated coupling coefficients
follow by the same substitutions in
Eqs.~\eqref{eq:pd-Xa-transport-coefficients}
and \eqref{eq:pd-ca0}--\eqref{eq:pd-ga2}.

To compare with Zhao's separating matrix~\cite{Zhao:2026gen},
let $\mathsf X_{\mathrm Z}$ denote the matrix in Eq.~(17)
of that reference, with the de Sitter
continuation of Sec.~IV. Its coordinates and mode labels obey
$t_{\mathrm Z}=t$, $\phi_{\mathrm Z}=\Xi\phi$,
$\omega_{\mathrm Z}=\omega$, and $m_{\mathrm Z}=m/\Xi$, so its
mode combinations $K,U$ coincide with our $K,W$.
Its parameters are related to ours by
$\kappa_{\mathrm Z}=\nu$ and
$\eta_{\mathrm Z}=\kappa_\star^{-1}$; in particular, its radial and
angular factors are $r^2-a^2/\kappa_\star$ and
$p^2+a^2/\kappa_\star$, respectively. On the common construction
patch, define the comparison factor
\begin{equation}
 \mathsf Z\coloneqq 
 \frac{(F_r\mathsf I+{Q}{\sqrt{\mathcal Q}}\sigma_2)({Q}\sqrt{P}\mathsf I+iF_{\theta}\sigma_2)}
      {\sqrt{F_r^2-Q^2\mathcal Q}\sqrt{F_\theta^2+Q^2P}}
 =e^{\left(f(r)+i\vartheta(\theta)\right)\sigma_2},
 \qquad \det\mathsf Z=1,
 \label{eq:pd-KNdS-frame-factor}
\end{equation}
where
\begin{equation}
 f(r)\coloneqq \frac12\log\frac{F_r+{Q}{\sqrt{\mathcal Q}}}{F_r-{Q}{\sqrt{\mathcal Q}}},\qquad
 \vartheta(\theta)\coloneqq \arg({Q}\sqrt{P}+iF_{\theta}).
 \label{eq:pd-KNdS-frame-functions}
\end{equation}
The functions are real, and $\sqrt{P}>0$ permits a smooth choice of the
argument on the angular interior. Direct multiplication, using
Eqs.~\eqref{eq:pd-radial-factorization}--\eqref{eq:pd-Phi-factorization}, gives
\begin{equation}
 \mathsf X=-i{Q}\sqrt\nu\,\mathsf X_{\mathrm Z}\mathsf Z.
 \label{eq:pd-KNdS-X}
\end{equation}
The constant factor changes
$\det\mathsf X_{\mathrm Z}=-1/(\nu {Q}^2)$ to one, while $\mathsf Z$
is precisely a separated radial boost and angular rotation of the
form in Eq.~\eqref{eq:separation-residual-gauge}. Thus the present
matrix and Zhao's matrix realize the same local separation planes.

For comparison with the transformation in
Ref.~\cite{Yoshida:2026yly}, Appendix~A.3 of that reference
gives its explicit relation to Zhao's matrix through
a separated radial and angular
change of basis. The coordinate and Fourier normalizations
are related to ours by Eq.~\eqref{eq:bg-KNdS-limit}.

To compare with the separating matrix of Sarkar
et al.~\cite{Sarkar:2026xkq}, consider their Eq.~(15).
With matching square-root branches, the matrix preceding
the exponential factor coincides with
Eq.~\eqref{eq:pd-KNdS-X-explicit}; the remaining factor
$e^{\psi\sigma_2}$ is a separated change of basis of the
form in Eq.~\eqref{eq:separation-residual-gauge}.

These comparisons hold
on the common regular patch. Physical horizon conditions require
the separate field-normalization analysis; for KNdS, see Sec.~5 of
Ref.~\cite{Yoshida:2026yly}.

\section{Conicity, global identifications and the angular spectrum}
\label{sec:global-angular}
\label{sec:pd-global-angular}
The local separation obtained above leaves the global mode conditions and the domain of the angular operator to be specified.
For the Pleba\'nski--Demia\'nski geometry, these depend on the azimuthal period and the treatment of the NUT time shifts.
We determine the admissible Fourier labels, classify the angular endpoints in the Bonnor and Misner interpretations, and establish an eigenfunction expansion for each self-adjoint angular realization at fixed real mode parameters.

\subsection{Conical defects at the axes}
\label{sec:conicity}
We evaluate the cone angles at fixed $r$ with $\mathcal Q(r)>0$, assuming $\mathcal P(\theta)>0$ on $[0,\pi]$ and finite, nonzero $\Sigma$ and $\Omega^2$ near the two axes.
$t_N$ denotes the time coordinate $t$ in Eq.~\eqref{eq:bg-metric}, and we define
\begin{equation}
 \begin{aligned}
 t_N&\coloneqq t,&
 A_N(\theta)&\coloneqq a\sin^2\theta+2\ell(1-\cos\theta),\\
 B_N(r)&\coloneqq r^2+(a+\ell)^2,&
 \mathcal P_0&\coloneqq\mathcal P(0),\qquad
 \mathcal P_\pi\coloneqq\mathcal P(\pi).
 \end{aligned}
 \label{eq:global-axis-deltas}
\end{equation}
The stationary Killing field $\partial_{t_N}$ is timelike near both poles.
We define the transverse cone angles using its orbit-space metric, $h_{ij}=g_{ij}-g_{t_Ni}g_{t_Nj}/g_{t_Nt_N}$.
Since $B_N-aA_N=\Sigma$, the angular components are
\begin{equation}
 \begin{aligned}
 h_{\theta\theta}&=\frac{\Sigma}{\Omega^2\mathcal P},\\
 h_{\phi\phi}
 &=g_{\phi\phi}-\frac{g_{t_N\phi}^2}{g_{t_Nt_N}}
 =\frac{\Sigma\mathcal Q\mathcal P\sin^2\theta}
 {\Omega^2(\mathcal Q-a^2\mathcal P\sin^2\theta)}.
 \end{aligned}
 \label{eq:global-transverse-metric}
\end{equation}
Let $x_N=\theta$ and $x_S=\pi-\theta$.
Here $\Sigma_A$, $\Omega_A$ and $\mathcal P_A$ denote their values at $\theta_N=0$ or $\theta_S=\pi$, for $A=N,S$, respectively.
At either endpoint,
\begin{equation}
 h_{x_Ax_A}=\frac{\Sigma_A}{\Omega_A^2\mathcal P_A}+O(x_A^2),
 \qquad
 h_{\phi\phi}=\frac{\Sigma_A\mathcal P_A}{\Omega_A^2}x_A^2
                    +O(x_A^4).
 \label{eq:global-axis-metric}
\end{equation}
For azimuthal period $2\pi\mathcal C$, with $\mathcal C>0$, the limiting circumference-to-radius ratio is
\begin{equation}
 \lim_{x_A\to0}
 \frac{2\pi\mathcal C\sqrt{h_{\phi\phi}}}
      {x_A\sqrt{h_{x_Ax_A}}}
 =2\pi\mathcal C\mathcal P_A.
 \label{eq:global-circumference}
\end{equation}
The signed deficits are therefore
\begin{equation}
 \delta_N=2\pi(1-\mathcal C\mathcal P_0),\qquad
 \delta_S=2\pi(1-\mathcal C\mathcal P_\pi).
 \label{eq:global-deficits}
\end{equation}
A positive deficit is conventionally associated with a cosmic string, and a negative deficit, or conical excess, with a strut.
The north conical deficit vanishes for
\begin{equation}
 \mathcal C=\mathcal P_0^{-1},\qquad
 \rho\coloneqq\frac{\mathcal P_\pi}{\mathcal P_0}>0,\qquad
 \delta_N=0,\qquad\delta_S=2\pi(1-\rho).
 \label{eq:global-north-conicity}
\end{equation}
The south pole then has a deficit for $0<\rho<1$ and an excess for $\rho>1$. The alternative normalization is
\begin{equation}
 \mathcal C=\mathcal P_\pi^{-1},\qquad
 \delta_S=0,\qquad\delta_N=2\pi(1-\rho^{-1}),
 \label{eq:global-south-conicity}
\end{equation}
for which the north pole has a deficit for $\rho>1$ and an excess for $0<\rho<1$.
A single period removes both conical deficits if and only if $\mathcal P_0=\mathcal P_\pi$. For the background of Sec.~\ref{sec:geometry},
\begin{equation}
 \mathcal P_\pi-\mathcal P_0
 =4a\left(bM-b^2\ell(a^2-\ell^2+Q^2)
                       -\frac{2\Lambda\ell}{3}\right).
 \label{eq:global-conicity-difference}
\end{equation}

\subsection{Time coordinates and azimuthal identifications}
\label{sec:time-charts}
Let $\upsilon$ be a real constant and introduce
\begin{equation}
 t_\upsilon=t_N-\upsilon\phi,\qquad
 A_\upsilon=A_N-\upsilon,\qquad B_\upsilon=B_N-a\upsilon.
 \label{eq:global-time-charts}
\end{equation}
The combinations $dt_N-A_Nd\phi$ and $a\,dt_N-B_Nd\phi$ retain their form in these coordinates.
The north-adapted chart has $\upsilon=0$, while $\upsilon=4\ell$ gives the south-adapted chart
\begin{equation}
 \begin{aligned}
 t_S&=t_N-4\ell\phi,\\
 A_S&=A_N-4\ell=a\sin^2\theta-2\ell(1+\cos\theta),\\
 B_S&=B_N-4a\ell=r^2+(a-\ell)^2.
 \end{aligned}
 \label{eq:global-south-chart}
\end{equation}
For nonperiodic $t_\upsilon$, we impose the azimuthal identification
\begin{equation}
 (t_\upsilon,\phi)\sim(t_\upsilon,\phi+2\pi\mathcal C).
 \label{eq:global-azimuthal-identification}
\end{equation}
It is generated by
\begin{equation}
 \left.\partial_\phi\right|_{t_\upsilon}
 =\left.\partial_\phi\right|_{t_N}+\upsilon\partial_{t_N}.
 \label{eq:global-azimuthal-generator}
\end{equation}
These generators induce the same azimuthal vector field on the stationary orbit space, so the cone angles in Eq.~\eqref{eq:global-deficits} are independent of $\upsilon$.
Under $\phi\mapsto\phi+2\pi\mathcal C$ at fixed $t_\upsilon$, the shifts in $t_N$ and $t_S$ are
\begin{equation}
 \Delta t_N=2\pi\mathcal C\upsilon,\qquad
 \Delta t_S=2\pi\mathcal C(\upsilon-4\ell).
 \label{eq:global-time-holonomies}
\end{equation}
The choices $\upsilon=0$ and $\upsilon=4\ell$ eliminate the north and south time shifts, respectively.
All other values give time-shift singularities at both poles.
In the north chart the identification reads $(t_N,\phi)\sim(t_N+2\pi\mathcal C\upsilon,\phi+2\pi\mathcal C)$.
Thus $\mathcal C$ fixes the transverse cone angles and $\upsilon$ specifies the time shift in the azimuthal identification.

A Fourier mode has the equivalent representations
\begin{equation}
 e^{-i\omega t_\upsilon+i m_\upsilon\phi}
 =e^{-i\omega t_N+i m_N\phi}
 =e^{-i\omega t_S+i m_S\phi},\qquad
 m_N=m_\upsilon+\upsilon\omega,\quad m_S=m_N-4\ell\omega.
 \label{eq:global-mode-shift}
\end{equation}
Here $m_N$ is the mode parameter denoted by $m$ in the preceding sections.
Single-valuedness under Eq.~\eqref{eq:global-azimuthal-identification} requires $m_\upsilon=k_\phi/\mathcal C$ with $k_\phi\in\mathbb Z$ in the azimuthally invariant polar frame of Sec.~\ref{sec:background}.\footnote{The radiative variables are neutral under Maxwell potential gauge transformations $\bm A\mapsto\bm A+\bm d\Lambda_{\rm gauge}$. Magnetic-potential patching therefore introduces no additional phase in their azimuthal identification.}
The mode combinations in the local equations become
\begin{equation}
 \begin{aligned}
 m_N-A_N\omega&=m_S-A_S\omega=m_\upsilon-A_\upsilon\omega,\\
 l&=a m_N-(a+\ell)^2\omega
   =a m_S-(a-\ell)^2\omega\\
  &=\frac{a k_\phi}{\mathcal C}
       -\bigl((a+\ell)^2-a\upsilon\bigr)\omega,
 \qquad K=l-\omega r^2.
 \end{aligned}
 \label{eq:global-invariant-modes}
\end{equation}
The separated equations retain their form under these substitutions.

\subsection{Endpoint classification of the angular operator}
\label{sec:angular-completeness}
Fix real mode parameters and a spin sector $\varsigma=\pm1$.
Assume $|\mathcal M|>0$, $\nu>0$, and that the construction
conditions in Eq.~\eqref{eq:pd-construction-patch} hold
throughout the angular interior on the radial interval
under consideration. In addition, require
\begin{equation}
 \mathcal P>0,\qquad a_0+\kappa_\star p^2>0
 \quad\hbox{on }[0,\pi].
 \label{eq:ang-regularity-assumptions}
\end{equation}
Source nondegeneracy is not required here, since the explicit
separation identities extend through zeros of $Z_{\rm src}$
under these conditions, as discussed in
Sec.~\ref{sec:pd-X-construction}.
The angular expression in Eq.~\eqref{eq:pd-La} acts in
\begin{equation}
 \mathcal H_\theta
 =L^2\bigl((0,\pi),\mathcal P^{-1/2}d\theta;\mathbb C^2\bigr).
 \label{eq:ang-hilbert-space}
\end{equation}
Since $\gamma_1,\gamma_2,\eta$ are real, $\mathcal L$ is formally symmetric on compactly supported smooth doublets.

The endpoint singularities occur only in $\gamma_2$. Expanding Eq.~\eqref{eq:pd-ga2} gives
\begin{equation}
 \gamma_2=-\frac{\varepsilon_N\sqrt{\mathcal P_0}\kappa_N}{\theta}
                  +O(1),\qquad
 \gamma_2=\frac{\varepsilon_S\sqrt{\mathcal P_\pi}\kappa_S}{\pi-\theta}
                  +O(1),
 \label{eq:angular-residues}
\end{equation}
with
\begin{equation}
 \begin{aligned}
 \kappa_N&=\frac32+\varsigma\frac{m_N}{\mathcal P_0}
 =\frac32+\varsigma
       \frac{k_\phi/\mathcal C+\upsilon\omega}{\mathcal P_0},\\
 \kappa_S&=\frac32-\varsigma\frac{m_S}{\mathcal P_\pi}
 =\frac32-\varsigma
       \frac{k_\phi/\mathcal C-(4\ell-\upsilon)\omega}{\mathcal P_\pi},\\
 \beta_N&=|\kappa_N|,\qquad\beta_S=|\kappa_S|.
 \end{aligned}
 \label{eq:ang-endpoint-exponents}
\end{equation}
The signs distinguish the two possible endpoint behaviors of $\kappa_\star-a_2p^2$:
\begin{equation}
 \varepsilon_A=
 \begin{cases}
 +1,&\kappa_\star-a_2p_A^2>0,\\
 -1,&\kappa_\star-a_2p_A^2=0,
 \end{cases}
 \qquad p_N=\ell+a,\quad p_S=\ell-a.
 \label{eq:angular-endpoint-signs}
\end{equation}
For $\varepsilon_A=+1$, the residue comes from the terms proportional to $\partial_\theta\sqrt{\mathcal P}+\sqrt{\mathcal P}\cot\theta$ and $(l+\omega p^2)/\sqrt{P}$.
If $\varepsilon_A=-1$, the factorization in Eq.~\eqref{eq:pd-Phi-factorization} instead gives $F_\theta(\theta_A)=0$ and
\begin{equation}
 -2a\cos\theta_A\,\mathcal P(\theta_A)
 =-\frac{2a_2p_A}{\kappa_\star}(a_0+\kappa_\star p_A^2)
 =-\frac{2\nu p_A}{Q^2},\qquad
 \theta_N=0,\quad\theta_S=\pi.
 \label{eq:angular-endpoint-factorization}
\end{equation}
The terms with denominator $\kappa_\star-a_2p^2$ then reverse the residue, while $\gamma_1$ and $\eta$ remain bounded since $F_\theta=O(x_A^2)$.
Near each endpoint, the angular equation is
\begin{equation}
 \frac{dS}{dx_A}
 =\frac{\varepsilon_A\kappa_A\sigma_2}{x_A}S+O(1)S.
 \label{eq:angular-fuchsian}
\end{equation}
Its characteristic powers are $x_A^{\beta_A}$ and $x_A^{-\beta_A}$; $\varepsilon_A$ affects the residue eigenvectors but not the indices.
The endpoint weight has a finite positive limit.
Hence
\begin{equation}
 \begin{cases}
 \beta_A\ge\tfrac12 &: \text{limit point},\\
 0\le\beta_A<\tfrac12 &: \text{limit circle}.
 \end{cases}
 \label{eq:angular-weyl-classification}
\end{equation}
At equality the $x_A^{-1/2}$ branch is not square integrable.
Resonant logarithms do not change this threshold, and for $\kappa_A=0$ the endpoint coefficients are bounded.
The Weyl classification uses nonreal angular spectral parameter $\zeta$ at fixed real spacetime frequency $\omega$.
\subsubsection{Bonnor interpretation: nonperiodic time}
\label{sec:bonnor}
In the Bonnor interpretation, time is nonperiodic and any Misner strings are retained~\cite{Bonnor:1969}.
We use the identification in Eq.~\eqref{eq:global-azimuthal-identification}, with $\mathcal C>0$ and $\upsilon\in\mathbb R$.
At fixed $k_\phi\in\mathbb Z$ and real $\omega$, Eq.~\eqref{eq:ang-endpoint-exponents} gives
\begin{equation}
 \begin{aligned}
 N\text{ is limit circle}
 &\quad\Longleftrightarrow\quad
 -2<\varsigma\frac{k_\phi/\mathcal C+\upsilon\omega}{\mathcal P_0}<-1,\\
 S\text{ is limit circle}
 &\quad\Longleftrightarrow\quad
 1<\varsigma\frac{k_\phi/\mathcal C-(4\ell-\upsilon)\omega}
                    {\mathcal P_\pi}<2.
 \end{aligned}
 \label{eq:bonnor-general-lc}
\end{equation}
Equality at any of these bounds is limit point.

For $\upsilon=0$, the integer is $k_N\coloneqq\mathcal C m_N=k_\phi$; for $\upsilon=4\ell$, it is $k_S\coloneqq\mathcal C m_S=k_\phi$.
Combining these two choices of fixed time with the two conical normalizations gives the cases in Table~\ref{tab:nonperiodic-identifications}.

\begin{table}[htbp]
\centering
\caption{Endpoint indices for azimuthal identifications at fixed
$t_N$ or $t_S$, with nonperiodic time. The integer label in each row
is defined with respect to the indicated fixed time;
$\beta_A=|\kappa_A|$.}
\label{tab:nonperiodic-identifications}
\begingroup
\small
\renewcommand{\arraystretch}{1.6}
\begin{tabular}{@{}llll@{}}
\toprule
$\mathcal C$ & Fixed time & $\kappa_N$ & $\kappa_S$\\
\midrule
$\mathcal P_0^{-1}$ & $t_N$
& $\frac32+\varsigma k_N$
& $\frac32-\varsigma\frac{k_N\mathcal P_0-4\ell\omega}{\mathcal P_\pi}$\\
$\mathcal P_0^{-1}$ & $t_S$
& $\frac32+\varsigma\left(k_S+\frac{4\ell\omega}{\mathcal P_0}\right)$
& $\frac32-\frac{\varsigma k_S}{\rho}$\\
$\mathcal P_\pi^{-1}$ & $t_N$
& $\frac32+\varsigma\rho k_N$
& $\frac32-\varsigma\left(k_N-\frac{4\ell\omega}{\mathcal P_\pi}\right)$\\
$\mathcal P_\pi^{-1}$ & $t_S$
& $\frac32+\varsigma\left(\rho k_S+\frac{4\ell\omega}{\mathcal P_0}\right)$
& $\frac32-\varsigma k_S$\\
\bottomrule
\end{tabular}
\endgroup
\end{table}

For $\mathcal C=\mathcal P_0^{-1}$ and identification at fixed $t_N$, the first row gives $\beta_N=|3/2+\varsigma k_N|\ge1/2$.
The north endpoint is therefore limit point, while the south endpoint is limit circle precisely when 
\begin{equation}
 1<\varsigma\frac{k_N\mathcal P_0-4\ell\omega}{\mathcal P_\pi}<2.
 \label{eq:bonnor-lc-window}
\end{equation}
For $\ell\ne0$, this is the open frequency interval
\begin{equation}
 |\omega-\omega_c|<\frac{\mathcal P_\pi}{8|\ell|},\qquad
 \omega_c=\frac{k_N\mathcal P_0-\frac32\varsigma\mathcal P_\pi}
                   {4\ell}.
 \label{eq:bonnor-frequency-window}
\end{equation}
Both boundaries are limit point.
With this identification, the indices become frequency independent when $\ell=0$.

Identification at fixed $t_S$ with $\mathcal C=\mathcal P_\pi^{-1}$ has a limit-point south endpoint.
For $\ell\ne0$, either mixed choice can admit two limit-circle endpoints.

\subsubsection{Misner interpretation: periodic time}
\label{sec:misner}
\label{sec:endpoint-summary}
For $\ell\ne0$, the Misner interpretation uses both adapted charts, with $t_N=t_S+4\ell\phi$, and identifies time with a positive period $T_M$~\cite{Misner:1963}.
Single-valuedness of this transition as a map into the time circle requires
\begin{equation}
 w_M\coloneqq\frac{8\pi\ell\mathcal C}{T_M}
       \in\mathbb Z\setminus\{0\}.
 \label{eq:misner-winding}
\end{equation}
The standard choice $|w_M|=1$ gives $T_M=8\pi|\ell|\mathcal C$; higher $|w_M|$ gives a finite quotient of the time fiber. Since $T_M>0$, $w_M$ has the sign of $\ell$.
This patching removes the time-string singularities, while the conical deficits are given by Eq.~\eqref{eq:global-deficits}.

The local mode representations $e^{-i\omega t_N+i m_N\phi}$ and $e^{-i\omega t_S+i m_S\phi}$ obey $m_S=m_N-4\ell\omega$.
Their consistency and single-valuedness imply
\begin{equation}
 \begin{aligned}
 \omega&=\frac{2\pi n_t}{T_M},\qquad n_t\in\mathbb Z,\\
 k_N&\coloneqq\mathcal C m_N\in\mathbb Z,
 &k_S&\coloneqq\mathcal C m_S\in\mathbb Z,\\
 N_\omega&\coloneqq4\ell\mathcal C\omega
              =k_N-k_S=w_Mn_t.
 \end{aligned}
 \label{eq:misner-mode-lattice}
\end{equation}
For $|w_M|=1$, $(k_N,N_\omega)$ may be taken as independent integers.
The substitution into the local equations is
\begin{equation}
 \omega=\frac{N_\omega}{4\ell\mathcal C},\qquad
 k_S=k_N-N_\omega,\qquad
 l=\frac{a k_N}{\mathcal C}-(a+\ell)^2\omega.
 \label{eq:misner-mode-substitution}
\end{equation}
For general $w_M$, $N_\omega$ is restricted to $w_M\mathbb Z$.\footnote{The map $(k_N,n_t)\mapsto(l,\omega)$ has determinant $aw_M/(4\ell\mathcal C^2)\ne0$.
The allowed mode family therefore retains full real span and preserves the source recoverability used in Sec.~\ref{sec:coupling-construction}.}

Under $\phi\mapsto\phi+2\pi\mathcal C$ at fixed $t_S$, $t_N$ shifts by $8\pi\ell\mathcal C=w_MT_M$.
The two adapted descriptions therefore correspond to integer bases of the same identification lattice.
In an arbitrary local chart $t_\upsilon$, the identification reads $(t_\upsilon,\phi)\sim (t_\upsilon-2\pi\mathcal C\upsilon,\phi+2\pi\mathcal C)$, with $t_\upsilon$ periodic modulo $T_M$.
A pure azimuthal identification at fixed $t_\upsilon$ represents the same lattice if and only if
\begin{equation}
 \frac{2\pi\mathcal C\upsilon}{T_M}
 =\frac{\upsilon}{4\ell}w_M\in\mathbb Z.
 \label{eq:misner-general-chart}
\end{equation}
Under this condition, $\mathcal C m_\upsilon=k_N-\upsilon N_\omega/(4\ell)$ is integral for every allowed time harmonic.

For a fixed winding, the two conical normalizations give
\begin{equation}
 \begin{array}{c|cc}
  \mathcal C&\kappa_N&\kappa_S\\[2pt]\hline
  \mathcal P_0^{-1}
   &\frac32+\varsigma k_N
   &\frac32-\frac{\varsigma k_S}{\rho}\\[6pt]
  \mathcal P_\pi^{-1}
   &\frac32+\varsigma\rho k_N
   &\frac32-\varsigma k_S
 \end{array}
 \label{eq:misner-indices}
\end{equation}
The conically regular endpoint is limit point in either normalization.
For $\mathcal C=\mathcal P_0^{-1}$, the south index depends on the time harmonic through $k_S=k_N-w_Mn_t$ at fixed $k_N$.
For $\mathcal C=\mathcal P_\pi^{-1}$, the north index depends on $k_N=k_S+w_Mn_t$ at fixed $k_S$.
The bounded angular coefficients retain their explicit frequency dependence.

For the north normalization $\mathcal C=\mathcal P_0^{-1}$, the south endpoint is limit circle precisely when
\begin{equation}
 \rho<\varsigma k_S<2\rho,\qquad k_S=k_N-N_\omega.
 \label{eq:misner-lc-window}
\end{equation}
At fixed time harmonic and spin sector, the integer labels satisfying this inequality are classified as follows:
\begin{center}
\begin{tabular}{@{}lll@{}}
\toprule
South conicity & Range & South limit-circle modes\\
\midrule
No defect & $\rho=1$ & none\\
Deficit & $0<\rho\le\tfrac12$ & none\\
Deficit & $\tfrac12<\rho<1$ & $k_S=\varsigma$\\
Excess & $\rho>1$ & $k_S=\varsigma n$,\quad $\rho<n<2\rho$\\
\bottomrule
\end{tabular}
\end{center}
Here $n$ is a positive integer and $k_N=N_\omega+k_S$.
The strict integer bounds and their count are
\begin{equation}
 \begin{aligned}
 n_{\min}&=\lfloor\rho\rfloor+1,&
 n_{\max}&=\lceil2\rho\rceil-1,\\
 N_{\rm LC}&=\lceil2\rho\rceil-\lfloor\rho\rfloor-1.
 \end{aligned}
 \label{eq:misner-strut-count}
\end{equation}
For every conical excess, $\rho>1$, this set is nonempty because $\rho<\lfloor\rho\rfloor+1\le\rho+1<2\rho$.
Its size is $\rho+O(1)$ as $\rho\to\infty$.
At either boundary of Eq.~\eqref{eq:misner-lc-window}, $\beta_S=1/2$ and the endpoint is limit point.

For the alternative normalization $\mathcal C=\mathcal P_\pi^{-1}$, the south endpoint is limit point and the north limit-circle labels are
\begin{equation}
 k_N=-\varsigma n,\qquad
 \rho^{-1}<n<2\rho^{-1},\qquad n\in\mathbb N.
 \label{eq:misner-south-period-lc}
\end{equation}
The classification follows from the same table with $\rho$ replaced by $\rho^{-1}$ and the pole and label signs interchanged.
For either normalization, only finitely many angular labels are limit circle at fixed time harmonic.

For a fixed time harmonic, the angular representatives satisfy
\begin{equation}
 f_S(\psi)=e^{-iN_\omega\psi}f_N(\psi),\qquad
 \psi\coloneqq\phi/\mathcal C\sim\psi+2\pi.
 \label{eq:misner-time-transition}
\end{equation}
This scalar transition is the line-bundle factor associated with the time Fourier mode. It commutes with $\mathsf X$ in the azimuthally invariant polar frame.

\subsection{Self-adjointness and angular completeness}
\label{sec:angular-consequences}
\label{sec:pd-angular-spectral-results}
Let $D_{\min}$ be the closure of $\mathcal L$ on $C_c^\infty((0,\pi);\mathbb C^2)$ at fixed real mode parameters.
If both endpoints are limit point, no boundary conditions are needed and
\begin{equation}
 \beta_N\ge\tfrac12,\quad\beta_S\ge\tfrac12
 \quad\Longrightarrow\quad
 D_{\min}\text{ is self-adjoint with compact resolvent}.
 \label{eq:ang-essential-self-adjointness}
\end{equation}
If one endpoint is limit circle, a self-adjoint boundary
condition must be imposed there~\cite{Teschl:1998}.
If both are limit circle, self-adjoint conditions may be
imposed separately at each end or may couple the boundary
data at the two ends.

First consider a self-adjoint realization with separated
boundary conditions. At $\zeta=i$, choose solutions square
integrable at the respective endpoints and satisfying any
imposed boundary conditions. Using these solutions, we
construct the Green kernel as in Sec.~4.2 of
Ref.~\cite{Yoshida:2026yly}.
At a limit-point endpoint, the same estimates, using the
endpoint powers above, give a square-integrable kernel.
At a limit-circle endpoint, both independent solutions
are square integrable, so their products also give a
square-integrable kernel.
The resolvent is therefore Hilbert--Schmidt and compact.

For any other self-adjoint realization, the difference
between its resolvent at $\zeta=i$ and the one just constructed
maps into the two-dimensional solution space of
$(\mathcal L-i)S=0$. This difference therefore has finite rank,
so compactness holds for every self-adjoint realization,
including those with coupled endpoint conditions.
Thus each specified realization has real, discrete spectrum
and a complete orthonormal eigenbasis in $\mathcal H_\theta$.

For real $g$ smooth on the open angular interval, multiplication
by $\mathsf U_g$ is unitary in $\mathcal H_\theta$ and maps
$C_c^\infty$ onto itself.
Let $\mathcal L_{\mathrm{sa}}$ denote a chosen self-adjoint
realization of $\mathcal L$.
The conjugation in Eq.~\eqref{eq:angular-frame-equivalence}
maps the minimal and maximal operators and carries the chosen domain
to $\mathsf U_g\operatorname{Dom}(\mathcal L_{\mathrm{sa}})$.
It preserves the spectrum, endpoint classification and
compactness of the resolvent.

For Misner identifications, the angular problem is considered within each allowed time harmonic, with the transition law in Eq.~\eqref{eq:misner-time-transition}.
For each fixed $r$, the bounded endpoint extensions of $\mathsf X$ and $\mathsf X^{-1}$ established in Sec.~\ref{sec:transformation-properties} preserve angular square integrability.

To obtain the angular expansion, write the transformed fields as
\begin{equation}
 \begin{aligned}
 \psi_+&=f_1e_++g_2e_-,&
 \psi_-&=g_1e_++f_2e_-,\\
 \mathbf f&\coloneqq\binom{f_1}{f_2},&
 \mathbf g&\coloneqq\binom{g_1}{g_2}.
 \end{aligned}
 \label{eq:general-reordered-fields}
\end{equation}
The radiative system then takes the form
\begin{equation}
 \begin{aligned}
 \partial_x\mathbf f&=-(\mathcal L+\xi)\mathbf f+(c_0+c_3)\mathbf g,\\
 \partial_x\mathbf g&=(\mathcal L+\xi)\mathbf g+(c_0-c_3)\mathbf f,
 \qquad \partial_x=\sqrt{\mathcal Q(r)}\partial_r.
 \end{aligned}
 \label{eq:general-operator-separation}
\end{equation}
Let $I_r$ be an open radial interval within the construction
patch, and choose the domain of $\mathcal L_{\mathrm{sa}}$ independently
of $r\in I_r$.
Let $\{S_n\}$ be an orthonormal eigenbasis satisfying $\mathcal L_{\mathrm{sa}} S_n=\zeta_nS_n$, with multiplicities included.
Suppose a solution satisfies $\mathbf f,\mathbf g\in C^1(I_r;\mathcal H_\theta)$, with both fields continuous into $\operatorname{Dom}(\mathcal L_{\mathrm{sa}})$ in the graph norm.
Then
\begin{equation}
 \begin{aligned}
 \mathbf f(r,\cdot)&=\sum_nR_{1n}(r)S_n,&
 \mathbf g(r,\cdot)&=\sum_nR_{2n}(r)S_n,\\
 R_{1n}&=\langle S_n,\mathbf f\rangle_{\mathcal H_\theta},&
 R_{2n}&=\langle S_n,\mathbf g\rangle_{\mathcal H_\theta}.
 \end{aligned}
 \label{eq:pd-angular-expansion}
\end{equation}
\equationrefalias{eq:pd-angular-expansion}{eq:pd-solution-expansion}
The series converge in $\mathcal H_\theta$.
Self-adjointness gives $\langle S_n,\mathcal L_{\mathrm{sa}}\mathbf f\rangle =\zeta_n\langle S_n,\mathbf f\rangle$, and the radial regularity permits differentiation of the coefficients.
Projecting Eq.~\eqref{eq:general-operator-separation} onto $S_n$ therefore gives Eq.~\eqref{eq:pd-Xa-radial} for $(R_{1n},R_{2n})$ with $\zeta=\zeta_n$.


\section[Summary and Discussion]{Summary and Discussion}
\label{sec:summary}
We have formulated a constructive criterion for local separation
of a specified first-order chiral system
(Theorem~\ref{thm:complete-criterion}).
For recoverable, nondegenerate source families, the criterion
gives necessary and sufficient conditions within the specified
class of mode-independent $SL(2,\mathbb C)$ transformations.
It also shows how strongly separation constrains the relation
between the coupling matrix $\mathsf F$ and the Hermitian
coefficient matrix $\mathsf A$.
For each admissible oriented Gram decomposition, the sources
of $\mathsf F$ determine a candidate separating matrix
$\mathsf X$ up to separated changes of frame.
With the differential operator fixed, wherever
$\mathsf J_{\mathrm I}\ne0$, the transport identity then
requires $\mathsf A$ to coincide with the specific matrix
$\mathsf A_F$ computed from these sources
(Theorem~\ref{thm:hermitian-transport}).
The remaining frame freedom leaves $\mathsf A_F$ unchanged
and therefore cannot accommodate an arbitrary choice of
$\mathsf A$.
Matching $\mathsf A$ is necessary but not sufficient:
the other transport conditions in
Eq.~\eqref{eq:hermitian-transport-tests} must also hold.
These tests can be evaluated directly from the source frame,
before reconstructing $\mathsf X$.
Once they succeed, the same frame gives an algebraic
construction of the separating matrix.

For the coupled radiative Einstein--Maxwell system on the PD
background with acceleration, NUT charge, and a cosmological
constant, the matrix $\mathsf X$ in Eq.~\eqref{eq:pd-Xa}
explicitly realizes separation on the specified construction
patch.
The resulting radial and angular equations are
Eqs.~\eqref{eq:pd-Xa-radial} and \eqref{eq:pd-Xa-angular},
respectively.
We construct this matrix from a separated source frame
after determining a Gram decomposition and verifying
the transport conditions. Under the additional assumptions
of Appendix~\ref{app:pd-rank-comparison}, the Gram decomposition
and its two planes are unique. These assumptions concern
uniqueness, rather than the existence of the constructed
decomposition. Even when the planes are unique, their bases
retain the freedom of a separated radial boost and angular
rotation, as given in Eq.~\eqref{eq:separation-residual-gauge}.
Direct substitution also extends the separation
identities through zeros of the source determinant under
the remaining patch assumptions, without asserting uniqueness
at those zeros.

The explicit matrix $\mathsf X$ requires no quadratures.
Its Kerr--Newman limit reproduces the known integrating
factor up to normalization, while its relation to existing
Kerr--Newman--de Sitter transformations is given by
normalization and separated changes of frame
(Sec.~\ref{sec:pd-special-backgrounds}).

Global identifications and conicity determine the allowed Fourier labels and angular endpoint indices without changing the local separation identities (Sec.~\ref{sec:pd-global-angular}).
Under the stated angular regularity assumptions, every
specified self-adjoint angular realization at fixed real
mode parameters has compact resolvent and a complete
orthonormal eigenbasis
(Sec.~\ref{sec:pd-angular-spectral-results}).
Limit-circle endpoints require additional boundary data
to select that realization.
For existing solutions satisfying the stated domain and
regularity assumptions, the expansion in
Eq.~\eqref{eq:pd-solution-expansion} yields the separated
radial equations for its coefficients.

The next physical step is to relate these angular domains
and separated solutions to the original spacetime fields.
For Kerr--Newman--de Sitter, Weller et al.\ reconstructed
the radiative metric and electromagnetic potential
perturbations from gauge-invariant master variables and
established parity isospectrality~\cite{Weller:2026bfn}.
For PD, reconstruction is needed to relate an angular
realization to field regularity or to a model of a retained
string or strut. Perturbing the defect parameters would
enlarge that problem. In the Bonnor interpretation,
quasinormal mode calculations also require physical axis
and horizon conditions and continuation to complex frequency.
Misner-periodic time instead imposes real discrete Fourier
frequencies and changes the usual resonance formulation.

Once these boundary conditions and the map from initial data
to radiative variables are established, the separated PD
system provides a starting point for studying sourced response
and time evolution. Recent work on Kerr--Newman has examined
exceptional points and mode exchange~\cite{Arcos:2026eis},
pseudospectral sensitivity near coalescing
modes~\cite{Destounis:2026qne}, and excitation
factors~\cite{Zhao:2026nce}.
For Kerr--Newman--de Sitter, the contribution of an exceptional
point to the ringdown waveform has been computed from initial
data satisfying the linearized Einstein--Maxwell
constraints~\cite{Destounis:2026xvd}.
Mode excitation and mixing have also been studied in
parametrized coupled perturbation systems for nonrotating
black holes in modified gravity~\cite{Macedo:2026pev}.
These developments motivate investigating the corresponding
response questions for PD beyond the determination of
quasinormal frequencies.

A related direction is to compare the present criterion
with separation through matrix angular operators.
Such constructions have been developed for scalar gravitational
perturbations of singly spinning Myers--Perry black holes that
are invariant under rotations of the transverse
sphere~\cite{Santos:2026pbu}, and for massive spin-2 fields
on Kerr~\cite{Dias:2026zri}.
In these constructions, commuting angular operators organize
coupled components into joint eigenfunctions.
A comparison would require determining whether the relevant
equations admit a formulation satisfying our chiral and source
conditions, or whether the criterion must be extended.

Beyond gravitational perturbations,
Appendix~\ref{app:other-systems} verifies the adjugate chiral
form for a translation-reduced Dirac Hamiltonian and the
Fu--Kane model at zero chemical potential~\cite{Fu:2008Kane},
and gives a simple separable subclass of the Dirac example.
Their source spans have dimension at most three, so applying
the nondegenerate necessity and uniqueness results requires
an extension to lower-rank source families.
Such an extension must account for the freedom in completing
the radial and angular planes and its compatibility with
the transport conditions.
Other settings for comparison include the chiral continuum
model of twisted bilayer graphene~\cite{Tarnopolsky:2019Magic},
inhomogeneous Maxwell equations~\cite{Khan:2005Maxwell},
generalized Weierstrass representations~\cite{Konopelchenko:1998Surfaces},
and the Davey--Stewartson scattering
problem~\cite{Astala:2015Scattering}.
In each case, the reductions, source ranks, and adjugate
relations must be checked before applying the criterion;
a chiral grading alone does not establish separation.
Separately, methods for Dirac and Bogoliubov--de Gennes
spectral problems, including those in the chiral
Gross--Neveu model~\cite{Basar:2008Kink}, may help analyze
the angular ODE after separation has been established.
\section*{Acknowledgments}
D.Y. is supported by JSPS KAKENHI Grant No. JP26K07084. The authors gratefully acknowledge access to OpenAI tools through the ChatGPT for Academic Researchers program.

\appendix
\section[Generic uniqueness of the PD Gram decomposition]{Generic uniqueness of the PD Gram decomposition}
\label{app:pd-rank-comparison}
\label{app:separation-necessity}
\label{sec:matrix-conditions}

Section~\ref{sec:pd-source-structure} determines the constant
matrix $\mathsf K$ uniquely once $\mathsf H_0$ is assigned
entirely to the radial plane and $\mathsf H_1$ to the angular
plane. We show that every admissible decomposition has this
assignment under the following conditions
\begin{equation}
 \mu-bnr^2\not\equiv0,\qquad
 n-b\mu p^2\not\equiv0,\qquad
 \det\Gamma\ne0,
 \label{eq:separation-generic-domain}
\end{equation}
where the first two functions must not vanish identically
on their respective coordinate intervals.
The last condition is the source-basis assumption already
used in Sec.~\ref{sec:pd-source-structure}.

Using the Gram entries in
Eqs.~\eqref{eq:pd-hr}--\eqref{eq:separation-proof-mode-gram},
we obtain the following determinants of $3\times3$ submatrices
of $\partial_r\Gamma$ and $\partial_\theta\Gamma$:
\begin{align}
 \det\left(\partial_r
 \begin{pmatrix}
 h_r & x_r & y_r\\
 x_r & \mathcal Q^{-1} & -r^2\mathcal Q^{-1}\\
 y_r & -r^2\mathcal Q^{-1} & r^4\mathcal Q^{-1}
 \end{pmatrix}\right)
 &=-\frac{2(\mu-bnr^2)}{\mathcal{Q}^2},
 \label{eq:separation-proof-radial-derivative-minor}\\
 \det\left(\partial_\theta
 \begin{pmatrix}
 h_\theta & x_\theta & y_\theta\\
 x_\theta & -P^{-1} & -p^2P^{-1}\\
 y_\theta & -p^2P^{-1} & -p^4P^{-1}
 \end{pmatrix}\right)
 &=\frac{2(n-b\mu p^2)}{a\sin\theta\,\mathcal P^2}.
 \label{eq:separation-proof-derivative-minors}
\end{align}

Consider any admissible decomposition. Since every inner
product of $\mathsf H_1$ with a source is independent of $r$,
the inner products of its radial component with all radial
source components are constant in $r$.
If this component were nonzero, an $r$-dependent Lorentz
transformation within the radial plane could locally make
it a fixed vector.
Denote the fixed radial projection of $\mathsf H_1$ by $u_0$
and the transformed radial projections of the sources by $v_a(r)$.
Since $\langle u_0,v_a(r)\rangle$ is constant,
$\langle u_0,\partial_r v_a\rangle=0$.
Thus all $\partial_r v_a$ lie in the one-dimensional
subspace orthogonal to $u_0$.
Writing $\partial_r v_a=c_a v_\perp$, where $v_\perp$
spans this subspace, we obtain
\begin{equation}
 \frac{d}{dr}\langle v_a,v_b\rangle
 =c_a\langle v_\perp,v_b\rangle+\langle v_a,v_\perp\rangle c_b.
 \label{eq:separation-proof-rank-lemma}
\end{equation}
The angular contribution to $\Gamma$ is independent of $r$,
so the left-hand side gives the entries of $\partial_r\Gamma$.
On the right-hand side, each term is an outer product of
two vectors indexed by the sources and therefore has rank
at most one. Their sum consequently has rank at most two.
However, wherever the determinant in
Eq.~\eqref{eq:separation-proof-radial-derivative-minor}
is nonzero, $\partial_r\Gamma$ has a $3\times3$ submatrix
with nonzero determinant and hence has rank at least three.
This contradiction rules out a nonzero radial component
of $\mathsf H_1$ at such points.
Interchanging the radial and angular roles and using
Eq.~\eqref{eq:separation-proof-derivative-minors}
similarly rules out a nonzero angular component of
$\mathsf H_0$ wherever that determinant is nonzero.
The assumptions ensure that the zeros of the numerators
are isolated, so both conclusions extend across them
by smoothness.

The assignment of $\mathsf H_0$ to the radial plane and
$\mathsf H_1$ to the angular plane fixes the corresponding
rows and columns of both Gram matrices.
The remaining freedom is confined to the
$(\mathsf T_l,\mathsf T_\omega)$ block and consists only
of adding a constant symmetric matrix $\mathsf K$ to
the radial part and subtracting it from the angular part.
Every admissible decomposition therefore has the form
in Eqs.~\eqref{eq:pd-radial-gram} and
\eqref{eq:separation-proof-gram-blocks}.
The rank conditions in Sec.~\ref{sec:pd-source-structure}
then fix $\mathsf K$ to Eq.~\eqref{eq:separation-gram-constant}.
This proves uniqueness of the Gram decomposition and,
by Eq.~\eqref{eq:general-source-frame}, of its two planes.
The matrix $\mathsf J$ is consequently fixed up to the
choice of radial orientation.

The first two conditions concern uniqueness only.
For $b\ne0$ they follow from $|\mathcal M|>0$, whereas
for $b=0$ they require $\mu n\ne0$.
In particular, the argument does not establish uniqueness
on the Kerr--Newman branch $b=n=0$, although the separating
matrix constructed in the main text remains available.

\section{Explicit chiral forms outside gravitational perturbation theory}
\label{app:other-systems}
The chiral structure studied in
Sec.~\ref{sec:general-separation-theorem} also appears
outside gravitational perturbation theory.
We illustrate this with a translation-reduced Dirac equation
and the Fu--Kane Hamiltonian at zero chemical potential,
writing both systems in the form
\[
 \mathscr D y_+=\mathsf Fy_-,\qquad
 \overline{\mathscr D}y_-=\mathsf F^\#y_+,
 \qquad \mathscr D\coloneqq \partial_x-i\partial_y,
\]
with $\mathsf A=0$. They establish a relation at the level of the differential
equations. Their source ranks must be examined separately before
applying the nondegenerate Gram construction. We set $\hbar=c=1$
in the Dirac example; the velocity in the second example is kept explicit.

\subsection{A translation-reduced Dirac Hamiltonian}
Let a four-component Dirac field obey $i\partial_t\Psi=H_D\Psi$, with
\begin{equation}
 H_D\coloneqq \boldsymbol\alpha\cdot(-i\boldsymbol\nabla-e\boldsymbol{\mathcal A})
       +\beta M(x,y)+V(x,y)\mathsf I_4,\qquad
 \boldsymbol{\mathcal A}\coloneqq (0,0,\mathcal A_z(x,y)),
 \label{eq:other-Dirac-H}
\end{equation}
where $M,V,\mathcal A_z$ are real and
$\alpha_j\coloneqq \begin{psmallmatrix}0&\sigma_j\\\sigma_j&0\end{psmallmatrix}$,
$\beta\coloneqq \operatorname{diag}(\mathsf I,-\mathsf I)$.
The function $M$ is a scalar mass background, so constant mass is included.
With the Fourier ansatz
$\Psi=e^{-i\omega t+ikz}(\phi_1,\phi_2,\chi_1,\chi_2)^T$,
the four component equations are
\begin{align}
 (W-M)\phi_1&=K_z\chi_1-i\mathscr D\chi_2,&
 (W-M)\phi_2&=-i\overline{\mathscr D}\chi_1-K_z\chi_2,\nonumber\\
 (W+M)\chi_1&=K_z\phi_1-i\mathscr D\phi_2,&
 (W+M)\chi_2&=-i\overline{\mathscr D}\phi_1-K_z\phi_2,
 \label{eq:other-Dirac-components}
\end{align}
where $W\coloneqq \omega-V$ and $K_z\coloneqq k-e\mathcal A_z$.
Define $y_+\coloneqq (\chi_2,\phi_2)^T$ and $y_-\coloneqq (\phi_1,\chi_1)^T$.
Solving the first and third equations for $\mathscr D y_+$, and the
fourth and second for $\overline{\mathscr D}y_-$, gives
\begin{equation}
 \mathsf F_D=i\begin{pmatrix}W-M&-K_z\\-K_z&W+M\end{pmatrix}
     =i(W\mathsf I-M\sigma_3-K_z\sigma_1),\qquad
 \mathsf G_D=i\begin{pmatrix}W+M&K_z\\K_z&W-M\end{pmatrix}=\mathsf F_D^\#.
 \label{eq:other-Dirac-F}
\end{equation}

For fixed backgrounds the mode family has the source presentation
\begin{equation}
 \mathsf F_D=i\omega \mathsf I-ik\sigma_1
        +i(-V\mathsf I-M\sigma_3+e\mathcal A_z\sigma_1).
 \label{eq:other-Dirac-source}
\end{equation}
All its Hermitian sources lie in
$\operatorname{span}_{\mathbb R}\{\mathsf I,\sigma_1,\sigma_3\}$, so their
pointwise span has dimension at most three.

A simple separable subclass has $V=V(x)$, $M=M(x)$ and
$\mathcal A_z=\mathcal A_z(y)$. Then $\mathsf X=\mathsf I$, $\mathsf T=\sigma_3$ and
$\xi=\eta=0$ satisfy the separation identities directly, with
$c_0=iW(x)$, $c_3=-iM(x)$, $c_1=-iK_z(y)$ and $c_2=0$.
The equations are
\begin{equation}
 \begin{pmatrix}-i\partial_y&iK_z\\-iK_z&i\partial_y\end{pmatrix}S
       =\zeta S,\qquad
 \partial_xR=
 \begin{pmatrix}-\zeta&i(W-M)\\
                 i(W+M)&\zeta\end{pmatrix}R.
 \label{eq:other-Dirac-separated}
\end{equation}
This is an application of Theorem~\ref{thm:chiral-separation}, whose
sufficiency does not require four independent sources. Arbitrary
two-variable backgrounds in Eq.~\eqref{eq:other-Dirac-H} remain subject
to a separation test; the component rearrangement alone does not
establish it.

\subsection{The Fu--Kane Hamiltonian at zero chemical potential}
In the Nambu convention of Ref.~\cite{Fu:2008Kane}, the single-particle
operator, written in the order $(u_1,u_2,v_1,v_2)^T$, is
\begin{equation}
 H_{\rm FK}\coloneqq -iv\tau_z(\sigma_1\partial_x+\sigma_2\partial_y)
       -\mu_{\rm c}\tau_z+\Delta_1\tau_x+\Delta_2\tau_y,
 \qquad \Delta\coloneqq \Delta_1+i\Delta_2.
 \label{eq:other-FK-H}
\end{equation}
Here $v>0$ is constant, $\tau_j$ act on the particle--hole blocks,
and $\sigma_j$ on spin; tensor products with identity matrices are
implicit. The real chemical potential $\mu_{\rm c}$ is distinct from
the PD parameter $\mu$. The upper and lower off-diagonal blocks of
$H_{\rm FK}$ are $\Delta^*\mathsf I$ and $\Delta \mathsf I$.
For $H_{\rm FK}\Psi=E\Psi$, set
$y_+\coloneqq (u_2,v_2)^T$ and $y_-\coloneqq (v_1,u_1)^T$. Component rearrangement gives
\begin{equation}
 \mathscr Dy_+=\mathsf F_{\mu_{\rm c}}y_-,\qquad
 \overline{\mathscr D}y_-=\mathsf G_{\mu_{\rm c}}y_+,
 \label{eq:other-FK-pair}
\end{equation}
where
\begin{equation}
 \mathsf F_{\mu_{\rm c}}=\frac{i}{v}
 \begin{pmatrix}-\Delta^*&E+\mu_{\rm c}\\\mu_{\rm c}-E&\Delta\end{pmatrix},
 \qquad
 \mathsf G_{\mu_{\rm c}}=\frac{i}{v}
 \begin{pmatrix}\Delta&\mu_{\rm c}-E\\E+\mu_{\rm c}&-\Delta^*\end{pmatrix}.
 \label{eq:other-FK-FG}
\end{equation}
Their adjugate defect is
\begin{equation}
 \mathsf G_{\mu_{\rm c}}-\mathsf F_{\mu_{\rm c}}^\#
   =\frac{2i\mu_{\rm c}}{v}\sigma_1.
 \label{eq:other-FK-defect}
\end{equation}
Thus at $\mu_{\rm c}=0$ this is exactly the chiral pair with $\mathsf A=0$ and
\begin{equation}
 \mathsf F_0=-\frac{\Delta_2}{v}\mathsf I
      -\frac{i\Delta_1}{v}\sigma_3-\frac{E}{v}\sigma_2.
 \label{eq:other-FK-Pauli}
\end{equation}
For fixed $\Delta(x,y)$ its Hermitian source span again has dimension
at most three. A vortex or another gap profile therefore needs a
separate construction of $\mathsf X$; the chiral form does not prove its
separation. For nonzero $\mu_{\rm c}$ the displayed rearrangement
fails the adjugate test.

\bibliographystyle{JHEP}
\bibliography{refs}

\end{document}